\documentclass[usenatbib]{emulateapj}
\usepackage{graphicx}
\usepackage[flushleft]{threeparttable}
\usepackage[usenames,dvipsnames,svgnames,table]{xcolor}
\usepackage{hyperref}
\definecolor{darkblue}{rgb}{0.0,0.0,0.3}
\hypersetup{colorlinks,breaklinks,
            linkcolor=darkblue,urlcolor=darkblue,
            anchorcolor=darkblue,citecolor=darkblue}
\usepackage{amsmath,amssymb}
\usepackage{amsmath}

\begin{document}

\def\etal{et al.\ \rm}
\def\ba{\begin{eqnarray}}
\def\ea{\end{eqnarray}}
\def\etal{et al.\ \rm}
\def\Fdw{F_{\rm dw}}
\def\Tex{T_{\rm ex}}
\def\Fdis{F_{\rm dw,dis}}
\def\Fnu{F_\nu}
\def\FJ{F_J}

\newcommand\cmtrr[1]{{\color{red}[RR: #1]}}

%%%%%%%%%%%%%%%%%%%%%%%%%%%%%%%%%%%%%%%%%%%%%%%%%%%%%%%%%%%

\title{Generalized Similarity for Accretion/Decretion Disks}

\author{Roman R. Rafikov\altaffilmark{1}}
\altaffiltext{1}{Institute for Advanced Study, 1 Einstein Drive, Princeton NJ 08540; 
rrr@ias.edu}

%%%%%%%%%%%%%%%%%%%%%%%%%%%%%%%%%%%%%%%%%%%%%%%%%%%%%%%%%%%

\begin{abstract}
Decretion (or external) disks are gas disks freely expanding to large radii due to their internal stresses. They are expected to naturally arise in tidal disruption events, around Be stars, in mass-losing post main sequence binaries, as a result of supernova fallback, etc. Their evolution is theoretically understood in two regimes: when the central object does not exert torque on the disk (a standard assumption for conventional accretion disks) or when no mass inflow (or outflow) occurs at the disk center. However, many astrophysical objects --- circumbinary disks, Be stars, neutron stars accreting in a propeller regime, etc. --- feature non-zero torque {\it simultaneously} with the non-zero accretion (or ejection of mass) at the disk center. We provide a general description for the evolution of such disks (both linear and non-linear) in the self-similar regime, to which the disk should asymptotically converge with time. We identify a similarity parameter $\lambda$, which is uniquely related to the degree, to which the central mass accretion is suppressed by the non-zero central torque. The known decretion disk solutions correspond to the two discrete values of $\lambda$, while our new solutions cover a {\it continuum} of its physically allowed values, corresponding to either accretion or mass ejection by the central object. A direct relationship between $\lambda$ and central $\dot M$ and torque is also established. We describe the time evolution of the various disk characteristics for different $\lambda$, and show that the observable properties (spectrum and luminosity evolution) of the decretion disks are in general different from the standard accretion disks with no central torque.
\end{abstract}


%%%%%%%%%%%%%%%%%%%%%%%%%%%%%%%%%%%%%%%%%%%%%%%%%%%%%%%%%%%

%%%%%%%%%%%%%%%%%%%%%%%%%%%%%%%%%%%%%%%%%%%%%%%%%%%%%%%%%%%
%%%%%%%%%%%%%%%%%%%%%%%%%%%%%%%%%%%%%%%%%%%%%%%%%%%%%%%%%%%

\section{Introduction}  
\label{sect:intro}

%%%%%%%%%%%%%%%%%%%%%%%%%%%%%%%%%%%%%%%%%%%%%%%%%%%%%%%%%%%

Astrophysical accretion disks evolve under the action of  internal stresses, which transport angular momentum outwards while most of the mass accretes onto the central object \citep{shakura_1973,PringlePhD, lynden-bell_1974}. Excess angular momentum shed by the accreted matter gets eventually transferred to the fluid at the outer edge of the disk, causing its outward {\it spreading}. Thus, while the inner parts of the disk get accreted, some gas must also move out to conserve angular momentum globally.

In many systems, such as X-ray binaries or cataclysmic variables, the outward expansion of the disk (fed with mass by the companion) gets eventually stopped by the torque exerted on the outer edge of the disk by the non-axisymmetric potential of the companion \citep{LinPap}. As a result, the disk attains a steady state, in which the excess angular momentum lost by the gas accreted by the central object gets passed to the binary orbit via the tidal disk-companion coupling.

On the other hand, there are many astrophysically relevant situations, in which the disk is not affected by external torques (at least temporarily). Is this case the disk keeps expanding while losing mass to central accretion. Such freely expanding disks are known as {\it decretion} or {\it external} disks \citep{Pringle} and are the subject of this study.

Decretion disks readily form when gas is deposited in orbit close to central object. This situation is natural for the tidal disruption events \citep{Komossa,Gezari}, in which the disk is assembled out of stellar debris close to a supermassive black hole \citep{Cannizzo,Shen}. Another important example is the disks of rapidly spinning Be stars \citep{Okazaki,Bes}, which have a long and diverse history of observations. Supernova fallback disks, expected to form out of the low angular momentum material not energetic enough to become unbound from the supernova remnant \citep{Michel81,Michel88}, should also evolve as decretion disks. The disk in which the planetary system around the millisecond pulsar PSR 1257 + 12 \citep{Pulsar} must have originated, has also likely undergone viscous expansion prior to planet formation  \citep{Phinney}, regardless of its origin. Post-main sequence evolution in stellar binaries often results in circumbinary disks fed by the mass outflow through the L2 Lagrange point \citep{Blundell,Dermine,Antoniadis}. This disk must also  subsequently evolve as a decretion disk torqued by the non-axisymmetric gravitational potential of the binary. 

Given their importance in astrophysics, it is not surprising that characterization of the decretion disk properties has a long history. In their pioneering work \citet{lynden-bell_1974} derived solutions for the viscous disk evolution assuming that the kinematic viscosity $\nu$ is independent of the surface density $\Sigma$. They explicitly demonstrated that in the long run the disk structure tends to evolve in a {\it self-similar fashion} independent of the {\it initial conditions}, i.e. the starting radial distribution of mass in the disk. Transition to this behavior naturally occurs when the disk expands far beyond the characteristic radius, at which most of its mass was concentrated initially. Analytical similarity solutions of \citet{lynden-bell_1974} have been subsequently generalized by \citet{lyubarskij_1987} to the nonlinear problem arising when the viscosity in the disk is an explicit function of $\Sigma$. 

On the other hand, the actual form of the long-term similarity solution does depend on the inner {\it boundary conditions} --- the rate of mass accretion and the torque applied to the disk at its center. \citet{lynden-bell_1974} and \citet{lyubarskij_1987} were able to find the similarity solutions for two important cases, which are common in nature. First is the disk with {\it zero (or very small) torque} at the center, akin to the accretion disk around a black hole \citep{shakura_1973}, which features a nonzero mass accretion rate $\dot M$ at the origin. Second is the disk with {\it zero mass inflow} at the center, $\dot M(r=0)=0$, in which the nonzero central torque must be present to completely suppresses accretion. This situation is thought to be typical e.g. for rapidly spinning, magnetized neutron stars that interact with the surrounding disk in the "propeller" regime \citep{Illarionov}. It was also thought, based on one-dimensional models \citep{Chang}, that $\dot M(r=0)=0$ is a natural boundary condition for the circumbinary disks around stellar binaries \citep{Alexander,Vartanyan} and supermassive black hole binaries \citep{ivanov_1999,Rafikov2013}, in which the binary torque would strongly suppress the gas inflow.

While these two orthogonal types of the central boundary conditions and their corresponding similarity solutions apply to many accreting systems, they certainly do not exhaust all astrophysically relevant possibilities. In fact, there is a number of objects accreting via the disk, in which {\it both} $\dot M$ {\it and} the torque do not vanish at the center, thus representing an {\it intermediate} situation when compared to the two known types of similarity solutions. In particular, magnetic field of the neutron stars in the propeller regime might present an imperfect obstacle to gas inflow. These stars would then accrete at some rate \citep{Arons}, as has been recently proposed to explain the accretion state transitions in Vela X-1 by \citet{Dorosh11} and in the ultraluminous X-ray source M82 X-2 \citep{M82} by \citet{Tsygankov}. Also, recent numerical work on circumbinary disks suggests that the torque produced by the  binary may not be efficient at suppressing gas accretion onto the binary \citep{MacFadyen2008,DOrazio2013,Farris2014}.

Moreover, in some systems the non-zero central torque is so strong that it results in $\dot M(r=0)<0$, i.e. {\it mass outflow} from the central object to the disk. Obvious example is given by the disks of the Be stars, which are fed by the gas shed from their rapidly spinning hosts \citep{Bes}. Circumbinary disks of post-main sequence binaries are also thought to be supplied by the gas lost from the binary, resulting in central injection of mass \citep{vanWinckel,deRuyter}. 

The goal of our present work is to explore the behavior of such systems, the evolution of which clearly cannot be captured by the two known decretion disk solutions \citep{lynden-bell_1974,Pringle}. Here we focus on deriving and analyzing the {\it self-similar} solutions, which describe the long-term evolution of the decretion disks. We do this for rather general class of the viscosity behaviors, both when $\nu$ is independent of $\Sigma$ (like in \citet{lynden-bell_1974}) and when it is an explicit function of $\Sigma$ (like in \citet{Pringle}). Moreover, our solutions naturally cover the possibility of both the inflow and the outflow at the disk center.

Our work is organized as follows.  In \S \ref{sect:basic} we convert the disk evolution equations to a form, which is particularly well suited for applying the self-similar ansatz (\S \ref{sect:nonlin}). After describing the previously known results (\S \ref{sect:zero_torque}-\ref{sect:zero_inflow}), we present our new self-similar solutions in \S \ref{sect:new}. We emphasize the important connection between the similarity parameter and the degree of the mass accretion suppression in \S \ref{sect:acc_supp}. Time evolution of the global characteristics of the decretion disks is covered in \S \ref{sect:time_evolve}, while their observables are described in \S \ref{sect:obs}. Finally, in \S \ref{sect:disc} we discuss the extraction of the self-similar disk parameters and provide comparison with the previous work (\S \ref{sect:compare}).

%%%%%%%%%%%%%%%%%%%%%%%%%%%%%%%%%%%%%%%%%%%%%%%%%%%%%%%%%%%
%%%%%%%%%%%%%%%%%%%%%%%%%%%%%%%%%%%%%%%%%%%%%%%%%%%%%%%%%%%

\section{Basic equations}  
\label{sect:basic}

%%%%%%%%%%%%%%%%%%%%%%%%%%%%%%%%%%%%%%%%%%%%%%%%%%%%%%%%%%%

We consider a thin disk in Keplerian rotation around a central mass $M_c$ (circumbinary disks orbit in the non-Newtonian potential, but far enough from binary the rotation profile converges to Keplerian). Local angular frequency of the disk fluid is $\Omega(r)= (GM_c/r^3)^{1/2}$ (we neglect radial pressure support), and $l(r)\equiv \Omega r^2$ is its specific angular momentum. We are interested in the azimuthally averaged, vertically integrated disk structure, so that all disk variables are functions of the radius $r$ and time $t$ only.

In this work we focus on the evolution driven by the internal (viscous) stresses alone. It is described by a one-dimensional (azimuthally-averaged) equation
\citep{lynden-bell_1974}
\ba
\frac{\partial \Sigma}{\partial t} & = & \frac{1}{2\pi r}
\frac{\partial \dot M}{\partial r},~~~~ \dot M = \left(\frac{d l}
{d r}\right)^{-1}\frac{\partial T_{r\phi}}{\partial r}.
\label{eq:evSigma}
\ea
Here $\dot M(r)$ is the local value of the mass accretion rate (defined to be positive for {\it inflow}), and $T_{r\phi}$ is the angular momentum flux due to the $r$-$\phi$ component of the {\it internal stress} in the disk (also equal to the total stress exerted by the disk interior to some $r$ on the outer disk, integrated over the circumference and height).

Equation (\ref{eq:evSigma}) assumes that there are no sources (or sinks) of the angular momentum and mass in the disk outside its very central part. In other words, external stress can be applied to the disk only at $r=0$ ($l=0$), giving rise to a non-zero value of $T_{r\phi}(r=0)$ as a boundary condition for the disk evolution. Similarly, $\dot M(r=0)$ is also in general non-zero.

Provided that stress is effected by some form of effective viscosity $\nu$, $T_{r\phi}$ is given by the {\it viscous} angular momentum flux $F_J$ \citep{filipov_1984,lyubarskij_1987,Rafikov2013}
\ba
T_{r\phi}=F_J\equiv -2\pi\nu\Sigma \frac{d\ln\Omega}{d\ln r}l=3\pi\alpha c_s^2\Sigma r^2,
\label{eq:F_J}
\ea
where $\nu$ is the kinematic viscosity usually expressed through the dimensionless parameter $\alpha$ and gas sound speed $c_s$ as $\nu=\alpha \Omega ^{-1}c_s^2$ \citep{shakura_1973}. Substituting $\FJ$ for $T_{r\phi}$ in equation (\ref{eq:dotM}) one arrives at the conventional form of the viscous evolution equation with $\Sigma(r,t)$ as the unknown function \citep{papaloizou_1995}. 

Equation (\ref{eq:evSigma}) can be recast in a particularly simple form by switching from surface density $\Sigma$ to the viscous angular momentum flux $\FJ$ and from $r$ to the specific angular momentum $l$ \citep{lynden-bell_1974,filipov_1984,lyubarskij_1987,Rafikov2013}:
\ba
\frac{\partial}{\partial t} \left( \frac{F_{J}}{D_{J}} \right) =
\frac{\partial^2\FJ}{\partial l^2},
\label{eq:evF_simple}
\ea
where
\ba
D_{J}\equiv -\nu r^2\frac{d\Omega}{dr}\frac{dl}{dr}=\frac{3}{4}\alpha c_s^2 l
\label{eq:D_J}
\ea
is the diffusion coefficient, which depends on $F_{J}$ if $\nu$ depends on $\Sigma$ (the second equality is for Keplerian $\Omega$). Once $\FJ$ is known from equation (\ref{eq:evF_simple}), the behavior of the surface density is given simply by 
\ba   
\Sigma=\frac{\FJ \Omega}{4\pi D_J}.
\label{eq:Sig}
\ea   
(specializing to the Keplerian rotation profile). It is also obvious that in these variables one can write
\ba
\dot M = \frac{\partial F_J}{\partial l},
\label{eq:dotM}
\ea
see equation (\ref{eq:evSigma}). In particular, a standard constant $\dot M$ accretion disk with zero torque at the center and extending to infinity \citep{shakura_1973} is described by a simple solution $F_J=\dot M l$.

%%%%%%%%%%%%%%%%%%%%%%%%%%%%%%%%%%%%%%%%%%%%%%%%%%%%%%%%%%%
%%%%%%%%%%%%%%%%%%%%%%%%%%%%%%%%%%%%%%%%%%%%%%%%%%%%%%%%%%%

\section{Self-similar ansatz}  
\label{sect:nonlin}

%%%%%%%%%%%%%%%%%%%%%%%%%%%%%%%%%%%%%%%%%%%%%%%%%%%%%%%%%%%

Now we consider a (generally nonlinear) problem, which arises when the diffusion coefficient $D_J$ is an explicit function of both $F_J$ and $l$. We will focus on the situation when $D_J$ has a power law dependence on both $F_J$ and $l$:
\ba
D_J=D_{J,0} F_J^d l^p,
\label{eq:DJpow}
\ea
where $D_{J,0}$, $d$, and $p$ are constants, which are set by the physics of the problem at hand (the behavior of $\nu$) using  equations (\ref{eq:F_J}), (\ref{eq:D_J}). This prescription is similar to \citet{Pringle}, who assumed viscosity to be a power law function of $\Sigma$ and $r$, namely $\nu\propto\Sigma^m r^n$. To simplify the comparison of the results, we provide the conversion between our variables and those of \citet{Pringle} in Appendix  \ref{sect:conversion}. In particular, equation (\ref{eq:conversion}) makes it clear that whenever $\nu$ depends on $ \Sigma$ (i.e. $m\neq 0$), one also has $D_J$ depending on $\FJ$ (i.e. $d\neq 0$). 

The values of $D_{J,0}$, $d$ and $p$ are determined by the model of the viscosity, namely its dependence on the disk properties. In general we will constrain $d$ to satisfy $0\le d<1$; values of $d>1$ result in viscous instability \citep{Lightman}. 

For our adopted $\alpha$-model the equation (\ref{eq:D_J}) shows that $D_J$ depends on the disk temperature $T$ via $c_s$. Externally illuminated disks typically have their temperature controlled only by the distance to the center, in which case both $T$ and $D_J$ are functions of $l$ only. This implies that $d=0$, i.e. that $D_J$ is independent of $\FJ$, making equation (\ref{eq:evF_simple}) linear. We will often refer to this type of situation as {\it linear} problem. It naturally emerges e.g. in irradiation-dominated regions of protoplanetary disks, see \citet{Vartanyan} for details. 

According to the equation (\ref{eq:D_J}), temperature scaling as $T(r)\propto r^{-k_T}$ results in $p=1-2k_T$ (so that $p<1$ for $T(r)$ decaying with distance). Centrally irradiated disks usually have $k_T$ close to $1/2$, so that $p\approx 0$. In particular, a passive disk model of \citet{Chiang} predicts $k_T=3/7$ for the optically thick part of the disk, resulting in $p=1/7$.

On the other hand, in disks heated by internal dissipation $T$ must be self-consistently determined by the local thermal balance and is in general a function of $\FJ$ (or $\Sigma$). As a result, $d\neq 0$ and equation  (\ref{eq:evF_simple}) becomes nonlinear (we refer to this situation as {\it nonlinear} problem). The effect of the details of the disk thermodynamics, namely the opacity behavior, on the parameters of the ansatz  (\ref{eq:DJpow}) has been previously explored by \citet{lyubarskij_1987}, \citet{Cannizzo}, \citet{Lipunova}. In particular, \citet{filipov_1984} and \citet{Rafikov2013} showed that for the gas-pressure dominated disk with the dominant free-free opacity $d=3/10$, $p=-4/5$. When the electron scattering opacity dominates $d=2/5$, $p=-6/5$. We will often use these values of $d$ and $p$ when describing our results.

Following \citet{filipov_1984}, \citet{lyubarskij_1987}, and \citet{filipov_1988} we will seek a self-similar solution of the equation (\ref{eq:evF_simple}) in the form
\ba
F_J(l,t)=F_0\varphi(t)f(\xi),~~~~\xi=\frac{l}{l_0\psi(t)},
\label{eq:ss_ansatz}
\ea
where $\xi$ is the similarity variable, $\varphi(t)$ and $\psi(t)$ are the dimensionless scaling functions and $l_0$ and $F_0$ are the dimensional scaling factors. 

Plugging in this ansatz together with (\ref{eq:DJpow}) into the equation (\ref{eq:evF_simple}) we find
\ba
\frac{D_0}{l_0^2}\varphi^{1+d}\psi^{p-2}
\xi^p f^d f^{\prime\prime}_{\xi\xi}=(1-d)\varphi^\prime_t \left(f-\frac{\psi^\prime_t}{\psi}\frac{\varphi}{\varphi^\prime_t}\xi f^\prime_\xi\right),
\label{eq:int_eq}
\ea
where $D_0\equiv D_{J,0}F_0^d l_0^p$ is the value of the viscosity coefficient set by the characteristic values of $F_J$ and $l$, and we use $s^\prime_z\equiv\partial s/\partial z$, etc. for any function $s$ and variable $z$.

Similarity of the solution obviously requires that
\ba
\psi=\varphi^\delta,
\label{eq:phi_psi}
\ea
where $\delta<0$ is a constant (the sign follows from the fact that the characteristic scale of the problem $l_0\psi$ must increase with time, while the amplitude $\propto\varphi$ decreases). We set a constant multiplier in the right hand side to unity because of the freedom in choosing both $F_0$ and $l_0$.

Equation (\ref{eq:int_eq}) then splits into two relations:
\ba
\varphi^\prime_t=-\frac{\varphi^{1+d+\delta (p-2)}}{(1-d)t_0},
\label{eq:time_eq}
\ea
which determines time evolution of the scale factor $\varphi$ (which is explored in detail in \S \ref{sect:time_evolve}), and 
\ba
\xi^p f^d f^{\prime\prime}_{\xi\xi}-\delta\xi f^\prime_{\xi}+f=0,
\label{eq:xi_eq}
\ea
which describes the overall spatial distribution of $F_J$. In equation (\ref{eq:time_eq}) we also defined a characteristic time $t_0\equiv l_0^2/D_0$.

Self-similar ansatz (\ref{eq:ss_ansatz}) together with equations (\ref{eq:F_J}), (\ref{eq:D_J}), (\ref{eq:DJpow}), and (\ref{eq:phi_psi}) allows us to write the instantaneous disk mass $M_d(t)$ as
\ba 
M_d &=& 2\pi\int\limits_0^\infty \Sigma(r) rdr=
\int\limits_0^\infty \frac{F_J}{D_J}dl=
\int\limits_0^\infty \frac{F_J^{1-d}l^{-p}}{D_{J,0}}dl
\nonumber\\
%%%%%
&=& \frac{F_0l_0}{D_0}\varphi^{1-d+\delta(1-p)}I_M,
\label{eq:M1}
\ea  
where we defined a new constant
\ba
I_M\equiv\int\limits_0^\infty f^{1-d}\xi^{-p}d\xi.
\label{eq:I_M}
\ea

The full angular momentum of the disk is given by 
\ba 
L_d = \frac{F_0l_0^2}{D_0}\varphi^{1-d+\delta(2-p)}I_L,~~~I_L\equiv\int\limits_0^\infty f^{1-d}\xi^{1-p}d\xi.
\label{eq:L1}
\ea  

Finally, with the help of equation (\ref{eq:dotM}) mass accretion rate becomes
\ba   
\dot M(l,t)=\frac{F_0}{l_0}\varphi^{1-\delta}f^\prime_\xi(\xi).
\label{eq:dotMxi}
\ea

By assumption, no torque is applied to the disk at its outer edge $\xi=\xi_{out}$ ($\xi_{out}$ may be equal to infinity), where $f(\xi_{out})=0$. It is easy to see that $\dot M$ must also vanish at this radius, meaning that $f^\prime_\xi(\xi_{out})=0$ according to equation (\ref{eq:dotMxi}). 

Mass conservation then implies that $\dot M(l=0,t)=-dM_d/dt$. Taking time derivative of equation (\ref{eq:M1}) and using equation (\ref{eq:time_eq}) we find that the requirement of mass conservation reduces to
\ba   
f^\prime_\xi(0)=\frac{1-d+\delta(1-p)}{1-d}I_M.
\label{eq:rel1}
\ea    
Analogously, global angular momentum conservation implies $F_J(l=0,t)=dL_d/dt$, which can be written as
\ba   
f(0)=-\frac{1-d+\delta(2-p)}{1-d}I_L.
\label{eq:rel2}
\ea    

It is important to note that equations (\ref{eq:rel1}) and (\ref{eq:rel2}) should be viewed as {\it consistency relations} rather than the boundary conditions for equation (\ref{eq:xi_eq}). Indeed, they directly follow 
from equation (\ref{eq:xi_eq}): multiplying it by $f^{-d}\xi^{-p}$ (or  $f^{-d}\xi^{1-p}$), integrating from $\xi=0$ to $\xi=\xi_{out}$, and using $f(\xi_{out})=0$, $f^\prime_\xi(\xi_{out})=0$ one immediately retrieves the relation (\ref{eq:rel1}) (or (\ref{eq:rel2})). Thus, these relations do not additionally constrain $f(\xi)$, but any solution of equation (\ref{eq:xi_eq}) must satisfy them. 

In the following we will use instead of $\delta$ a new {\it similarity parameter} $\lambda$ defined as
\ba
\lambda\equiv 1+\delta\frac{1-p}{1-d},~~\Longrightarrow ~~\delta=-\frac{(1-\lambda)(1-d)}{1-p}.
\label{eq:lambda}
\ea
With this new parameter equation (\ref{eq:xi_eq}) transforms to 
\ba
\xi^p f^d f^{\prime\prime}_{\xi\xi}-\frac{(\lambda-1)(1-d)}{1-p}\xi f^\prime_{\xi}+f=0,
\label{eq:xi_eq1}
\ea
subject to conditions
\ba   
f^\prime_\xi(0)=\lambda I_M,~~~~f(0)=-\left(\lambda+\frac{\lambda-1}{1-p}\right)I_L.
\label{eq:rels}
\ea    

Equation (\ref{eq:xi_eq1}) possesses an important symmetry property: if some function $g(\xi)$ is its solution, then a function $f(\xi)=k^{(p-2)/d}g(k\xi)$ also satisfies equations (\ref{eq:xi_eq1})-(\ref{eq:rels}). This rescaling also results in 
\ba    
I_M(f) &=& k^{1+(p-2)/d}I_M(g),
\label{eq:IM_change}\\
%%%%%
I_L(f) &=& k^{(p-2)/d}I_L(g).
\label{eq:IL_change}
\ea
This gauge freedom affects the choice of the dimensional parameters $F_0$ and $l_0$ characterizing the amplitude and scale of the density distribution in the disk. For a disk with fixed mass and angular momentum this follows immediately from equations (\ref{eq:M1}) and (\ref{eq:L1}). To fix this gauge dependence in what follows we constrain the solution $f$ of equation (\ref{eq:xi_eq1}) to satisfy the condition $I_M=1$. The value of $I_L$ then follows from equation (\ref{eq:IL_change}).

Another benefit of the scaling symmetry property of the  equation (\ref{eq:xi_eq1}) is that it allows lowering its order \citep{lyubarskij_1987}. However, this provides useful insights only in a couple of cases considered below.

%%%%%%%%%%%%%%%%%%%%%%%%%%%%%%%%%%%%%%%%%%%%%%%%%%%%%%%%%%%

\subsection{Previously known solutions: no central torque, $\lambda=(2-p)^{-1}$.}  
\label{sect:zero_torque}

%%%%%%%%%%%%%%%%%%%%%%%%%%%%%%%%%%%%%%%%%%%%%%%%%%%%%%%%%%%

The problem of the decretion disk evolution is known to admit two types of analytical solutions. One of them corresponds to the standard assumption commonly used in modelling accretion disks \citep{shakura_1973} --- that of zero (or very small) central torque, when $F_J(r=0)=0$ or $f(\xi=0)=0$ and the angular momentum of the disk is conserved, $L_d(t)=$ const. It is obvious that the zero central torque assumption must correspond to 
\ba    
\lambda=\lambda_0\equiv (2-p)^{-1},
\label{eq:lamLd}
\ea    
as this value of $\lambda$ naturally reduces the second constraint in (\ref{eq:rels}) to $f(0)=0$. 

\begin{figure}
\centering
\includegraphics[width=0.5\textwidth]{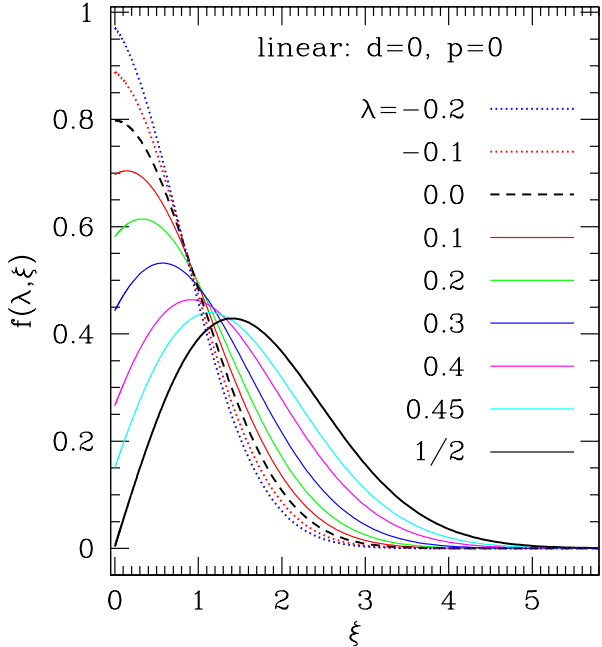}
\caption{Self-similar solutions for the function $f(\lambda,\xi)$ for several values of $\lambda$ labeled on the panels for individual curves. Disk with linear viscosity is assumed, i.e. $d=0$, and the diffusion coefficient $D_J$ is taken to be independent of radius (or $l$), i.e. $p=0$, as appropriate for a constant $\alpha$ disk with the midplane temperature profile $T\propto r^{-1/2}$. Both inflow (solid curves) and outflow (dotted curves) solutions are shown. Black solid and dashed curves correspond to solutions with zero central torque ($\lambda=(2-p)^{-1}$, see \S \ref{sect:zero_torque}, Eq. [\ref{eq:linLcons}]) and zero central mass flow ($\lambda=0$, see \S \ref{sect:zero_inflow}, Eq. [\ref{eq:linMcons}]), respectively.  
\label{fig:f_lin}}
\end{figure}

Similarity solution for the {\it linear} problem ($d=0$) with no central torque was derived in the pioneering study of \citet{lynden-bell_1974}:
\ba     
f(\lambda_0,\xi)=\frac{\xi}{2-p}\exp\left[-\frac{\xi^{2-p}}{(2-p)^2}\right],
\label{eq:linLcons}
\ea     
where we set the normalization to guarantee that $I_M=1$. This solution is shown by the black solid curve in Figure \ref{fig:f_lin}. 

%%%%%%%%%%%%%%%%%%%%%%%%%%%
%%%%%%%%%%%%%%%%%%%%%%%%%%%

%%%%%%%%%%%%%%%%%%%%%%%%%%%
\begin{figure}
\centering
\includegraphics[width=0.5\textwidth]{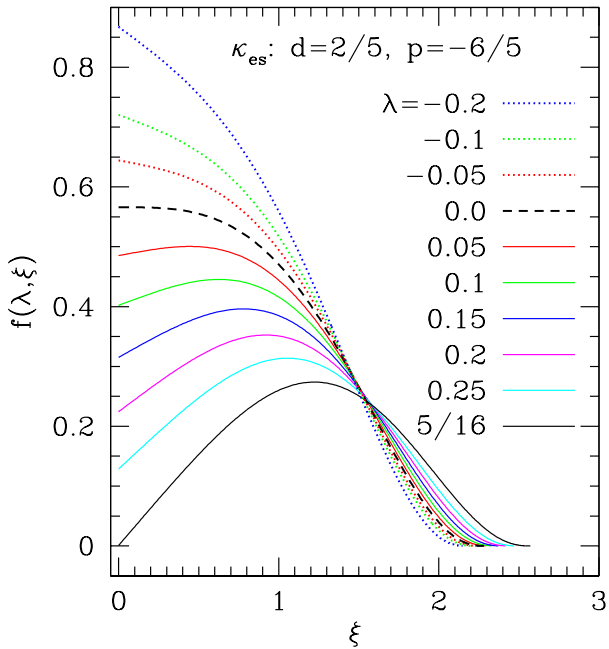}
\caption{Same as Figure \ref{fig:f_lin} but for a disk with opacity dominated by the electron scattering ($d=2/5$, $p=-6/5$), when disk evolution is a {\it non-linear} problem. Note the finite extent in $\xi$ of the solutions for different $\lambda$. Black solid and dashed curves correspond to solutions (\ref{eq:solLdconst}) and (\ref{eq:solMdconst}). 
\label{fig:f_es}}
\end{figure}
%%%%%%%%%%%%%%%%%%%%%%%%%%%

%%%%%%%%%%%%%%%%%%%%%%%%%%%
\begin{figure}
\centering
\includegraphics[width=0.5\textwidth]{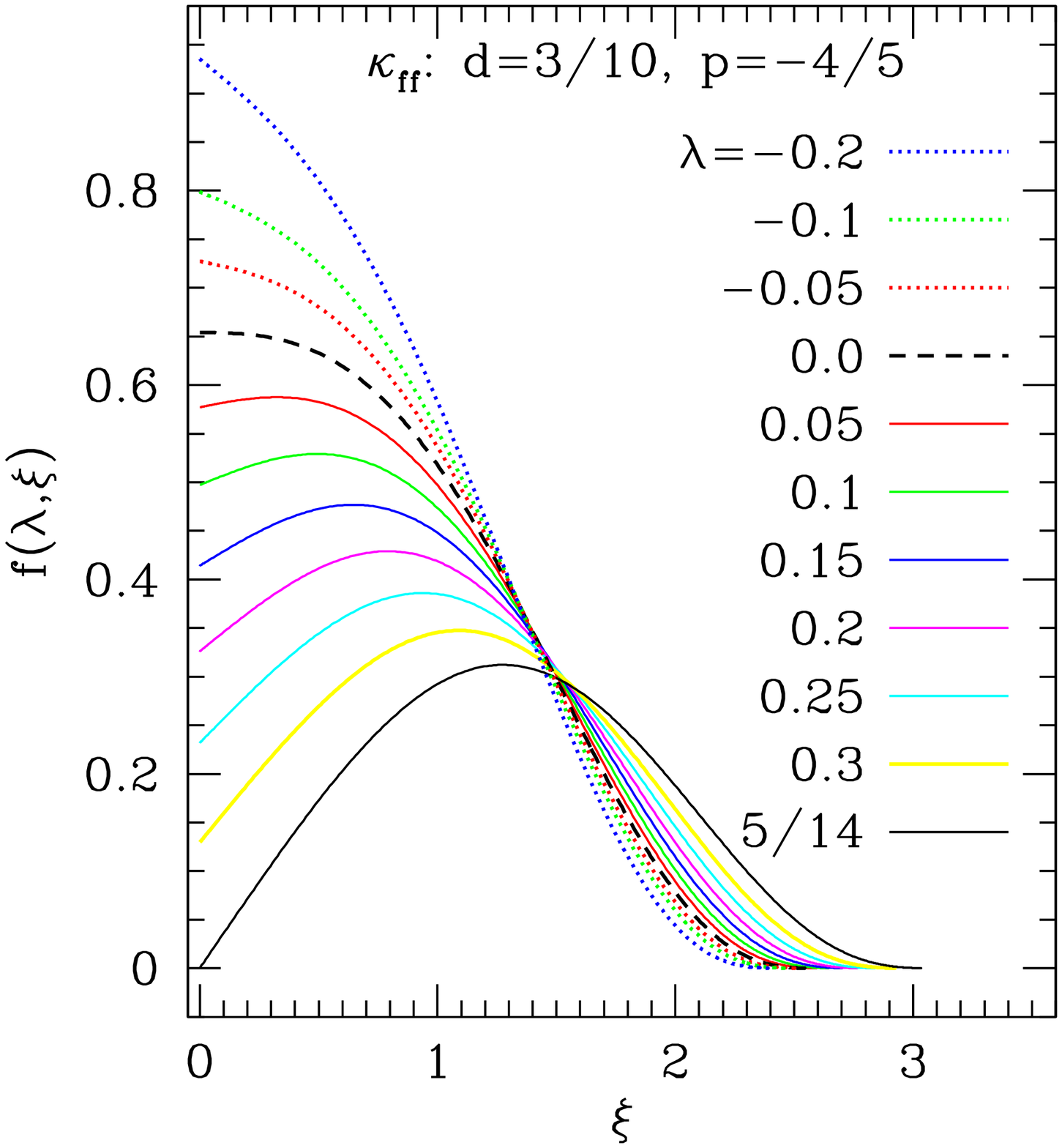}
\caption{Same as Figure \ref{fig:f_es} but for a disk with the free-free opacity ($d=3/10$, $p=-4/5$). 
\label{fig:f_ff}}
\end{figure}
%%%%%%%%%%%%%%%%%%%%%%%%%%%

In the nonlinear case ($d>0$) the analytical similarity solution conserving $L_d$ was derived by \citet{lyubarskij_1987}, using the results of \citet{BarenZel} in the field of gas filtration \citep{BarenBook}, and subsequently by \citet{Cannizzo}. It follows from equation (\ref{eq:xi_eq1}) that such solution has a form \citep{Pringle}
\ba
f(\xi)=(2-p)^{-1}\xi\left(1-c_1\xi^{2-p-d}\right)^{1/d},
\label{eq:solLdconst}
\ea    
with constant factor $c_1$ given by equation (\ref{eq:c1}) to guarantee $I_M=1$. This solution clearly satisfies the first consistency relation (\ref{eq:rels}). Its central mass accretion rate is non-zero and given by $\dot M(r=0,t)=(2-p)^{-1}(F_0/l_0)[\varphi(t)]^{1-\delta}$, see equation (\ref{eq:dotMxi}). The value of $I_L$ for this $\lambda$ is given by equation (\ref{eq:ILlam0}). The shape of this solution is illustrated in Figures \ref{fig:f_es} \& \ref{fig:f_ff} (black solid curve).

%%%%%%%%%%%%%%%%%%%%%%%%%%%%%%%%%%%%%%%%%%%%%%%%%%%%%%%%%%%

\subsection{Previously known solutions: no central inflow, $\lambda=0$.}  
\label{sect:zero_inflow}

%%%%%%%%%%%%%%%%%%%%%%%%%%%%%%%%%%%%%%%%%%%%%%%%%%%%%%%%%%%

Second previously known analytical self-similar solution corresponds to zero mass flux at the center, i.e. $\dot M(r=0)=0$ or $f^\prime_\xi(0)=0$. According to the relation (\ref{eq:rels}) this requires $\lambda=0$. In this case the total mass of the disk is conserved, $M_d(t)=$ const. 

Linear ($d=0$) similarity solutions of this kind were again obtained by \citet{lynden-bell_1974}:
\ba     
f(0,\xi)=c_2\exp\left[-\frac{\xi^{2-p}}{(1-p)(2-p)}\right],
\label{eq:linMcons}
\ea     
with $c_2$ given by equation (\ref{eq:c5}) to ensure $I_M=1$.

In the {\it nonlinear} case ($d>0$) the mass-conserving ($\lambda=0$) solution of the equation (\ref{eq:xi_eq1}) was derived by \citet{lyubarskij_1987}, using the results obtained by \citet{Zel} and \citet{Baren} in their studies of the nonlinear heat conduction and gas filtration \citep{BarenBook}. It reads (subject to the constraint $I_M=1$)
\ba
f(\xi) &=& c_3\left(1-c_4\xi^{2-p}\right)^{1/d},
\label{eq:solMdconst}
\ea  
in agreement with \citet{Pringle}. The constant factors $c_3$ and $c_4$ are given by equations (\ref{eq:c2}), (\ref{eq:c3}) and yield $I_M=1$. As the disk viscously spreads while preserving its mass, the central torque varies as $F_J(r=0,t)=c_3F_0\varphi(t)$.  The value of $I_L$ for this $\lambda$ is given by equation (\ref{eq:IL0}).
The shape of this solution is shown by the dashed black curve in Figures \ref{fig:f_es} \& \ref{fig:f_ff}.

%%%%%%%%%%%%%%%%%%%%%%%%%%%%%%%%%%%%%%%%%%%%%%%%%%%%%%%%%%%
%%%%%%%%%%%%%%%%%%%%%%%%%%%%%%%%%%%%%%%%%%%%%%%%%%%%%%%%%%%

\section{New similarity solutions}  
\label{sect:new}

%%%%%%%%%%%%%%%%%%%%%%%%%%%%%%%%%%%%%%%%%%%%%%%%%%%%%%%%%%%

\begin{figure}
\centering
\includegraphics[width=0.47\textwidth]{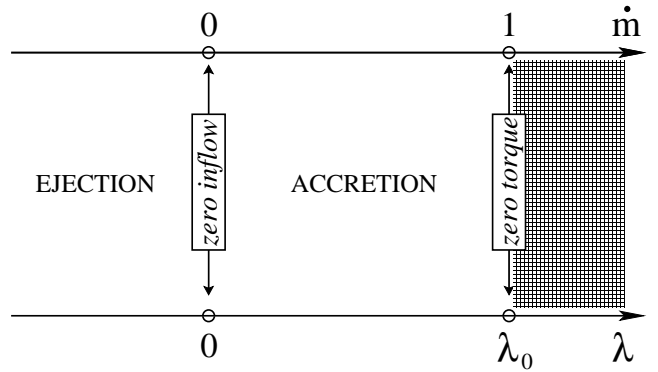}
\caption{Schematic illustration of the various possibilities (accretion or ejection of mass by the central object) related to different values of the similarity parameter $\lambda$ (lower axis). Shaded region to the right of $\lambda_0$ corresponds to $\FJ(r\to 0)<0$, which is unphysical for the stress model (\ref{eq:F_J}). The upper axis corresponds to $\dot m$ --- the degree, to which accretion by the central object is suppressed by the nonzero central torque compared to the case of zero central torque. The direct (but nonlinear) connection between $\dot m$ and $\lambda$ is established in \S \ref{sect:acc_supp}.
\label{fig:scheme}}
\end{figure}

To the best of our knowledge, equations (\ref{eq:linLcons})-(\ref{eq:solMdconst}) represent the only two self-similar decretion disk solutions that have been discussed in the literature. In the space of possible values of the similarity parameter $\lambda$ they cover just two discrete points, leaving the continuum of other values of $\lambda$ unaddressed. This is illustrated in Figure \ref{fig:scheme} that displays various possibilities related to different values of $\lambda$. Our goal here is to provide a description of the self-similar solutions for {\it all possible} values of $\lambda$ and to connect them to the physical properties of the specific systems. 

First of all, from the boundary conditions (\ref{eq:rels}) it is clear that unless
\ba
\lambda\ge \lambda_0=(2-p)^{-1}
\label{eq:phys_meaning}
\ea    
the central torque on the disk would become negative. This is not possible since in our setup $\FJ$ is directly related to $\Sigma$, meaning that $f(0)<0$ would imply $\Sigma(r=0)<0$. Thus, physically meaningful similarity solutions are possible only for the values of $\lambda$ satisfying the constraint (\ref{eq:phys_meaning}). 

Boundary conditions (\ref{eq:rels}) also make it clear that the negative values of $\lambda$ result in {\it mass outflow} at the center, because $\dot M(r=0)\propto \lambda$. This situation describes the mass injection by the central object into the disk, as expected e.g. in decretion disks of Be stars.

%%%%%%%%%%%%%%%%%%%%%%%%%%%%%%%%%%%%%%%%%%%%%%%%%%%%%%%%%%%

\subsection{Linear problem}  
\label{sect:lin}

%%%%%%%%%%%%%%%%%%%%%%%%%%%%%%%%%%%%%%%%%%%%%%%%%%%%%%%%%%%

We first describe similarity solution for the {\it linear} ($D_J$ independent of $\FJ$, $d=0$) viscous evolution problem, when the diffusion coefficient $D_J$ is a function of $l$ only. In Appendix \ref{sect:lin_details} we show that the general linear solution of the equation (\ref{eq:xi_eq1}) can be expressed analytically as  
\ba   
f(\lambda,\xi)=c_5 e^{-\kappa\xi^{2-p}}U(b-a,b,\kappa\xi^{2-p}),
\label{eq:lin_sol}
\ea    
where constant factors  $\kappa(\lambda)$, $a$, $b$, and $c_5(\lambda)$ are given by equations (\ref{eq:z}), (\ref{eq:ab}),  and (\ref{eq:c4}) respectively, and $U(c,q,t)$ is the Tricomi confluent hypergeometric function \citep{Abram}. This solution is plotted in Figure \ref{fig:f_lin} for different values of $\lambda$.

One can easily show that this solution naturally satisfies the constraints (\ref{eq:rels}) at the origin. Far from the origin $f(\xi)$ rapidly decays as 
\ba    
f(\xi)~\sim ~\xi^{\lambda(1-p)/(1-\lambda)}\exp\left(-\kappa\xi^{2-p}\right),~~~\xi\to\infty.
\label{eq:asympt}
\ea    
Thus, the solution for $\Sigma$-independent viscosity extends to infinity for any $\lambda$.

%%%%%%%%%%%%%%%%%%%%%%%%%%%
%%%%%%%%%%%%%%%%%%%%%%%%%%%

Using basic properties of the Tricomi function \citep{Abram} one can easily show that in the disk with zero torque the solution (\ref{eq:lin_sol}) reduces to equation (\ref{eq:linLcons}), while for $\lambda=0$ (no central inflow) one reproduces equation (\ref{eq:linMcons}).

A nice feature of the linear problem is that many of its properties can be written explicitly. For example, equations (\ref{eq:IL_lin}) and (\ref{eq:dotm_lin}) provide analytical expressions for the angular momentum integral $I_L$ and the degree $\dot m$ to which the central accretion is suppressed, see \S \ref{sect:acc_supp}. Behavior of these and some other characteristics of the linear solutions are shown in Figure \ref{fig:con_lin} as functions of $\lambda$.

%%%%%%%%%%%%%%%%%%%%%%%%%%%%%%%%%%%%%%%%%%%%%%%%%%%%%%%%%%%

\subsection{Nonlinear problem}  
\label{sect:nlin}

%%%%%%%%%%%%%%%%%%%%%%%%%%%%%%%%%%%%%%%%%%%%%%%%%%%%%%%%%%%

We next explore the case of the nonlinear viscosity, $d>0$. Restricting ourselves to the range (\ref{eq:phys_meaning}) we numerically solve equation (\ref{eq:xi_eq1}) for different values of $\lambda$ subject to the additional constraint $I_M=1$. Our results are shown in Figure \ref{fig:f_es} for $d=2/5$, $p=-6/5$ ($\kappa_{\rm es}$ regime) and Figure \ref{fig:f_ff} for $d=3/10$, $p=-4/5$ ($\kappa_{\rm ff}$ regime). In Figures \ref{fig:con_es}, \ref{fig:con_ff} we display the behavior of various characteristics of these nonlinear solutions --- $I_L$, $f_0$ (directly related to the amplitude of the central torque), etc. --- as functions of $\lambda$.

Note that all solutions of the nonlinear problem have $F_J$ and, consequently, $\Sigma$ vanishing at a finite radius. This is a characteristic feature of the nonlinear diffusion, in which the speed of signal propagation is limited, unlike the linear problem of \S \ref{sect:lin}, in which the (exponentially suppressed) tail of $\Sigma$ distribution extends out to infinity almost instantaneously.

Solutions for other values of $d$ and $p$ can be found analogously, by numerically solving equation (\ref{eq:xi_eq1}). We were unable to identify other obvious analytical solutions of this nonlinear equation, apart from the known results described in \S \ref{sect:zero_torque} \& \ref{sect:zero_inflow}. 

\begin{figure}
\centering
\includegraphics[width=0.5\textwidth]{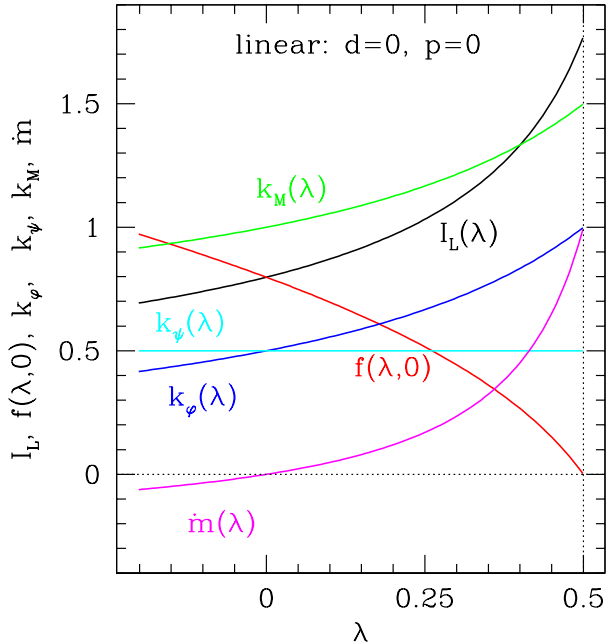}
\caption{Behavior of different characteristics of our self-similar solutions as functions of $\lambda$, for a disk with linear viscosity ($d=0$) and $p=0$ (see Fig. \ref{fig:f_lin}). Shown are the angular momentum integral $I_L$ defined by equation (\ref{eq:L1}), degree of suppression of the central mass accretion $\dot m$, $f(\xi=0)$ describing the strength of the central torque, as well as the time scaling exponents $k_\varphi$ (for $\varphi$ and $\FJ(0,t)$), $k_\psi$ (for $\psi$) and $k_M$ (for $\dot M(t)$), see equations (\ref{eq:k_varphi}), (\ref{eq:k_psi}),  and (\ref{eq:k_M}). 
\label{fig:con_lin}}
\end{figure}

\begin{figure}
\centering
\includegraphics[width=0.5\textwidth]{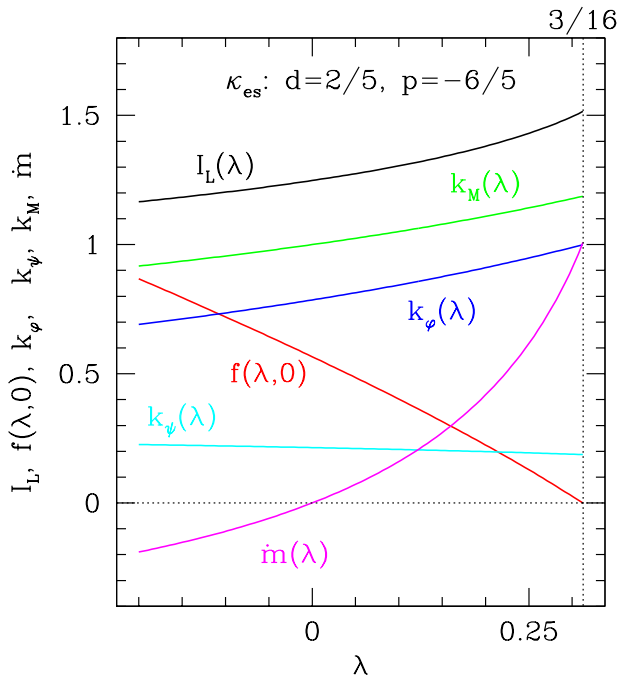}
\caption{Same as Figure \ref{fig:con_lin} but for the non-linear solutions in a disk with the electron scattering opacity ($d=2/5$, $p=-6/5$). 
\label{fig:con_es}}
\end{figure}

\begin{figure}
\centering
\includegraphics[width=0.5\textwidth]{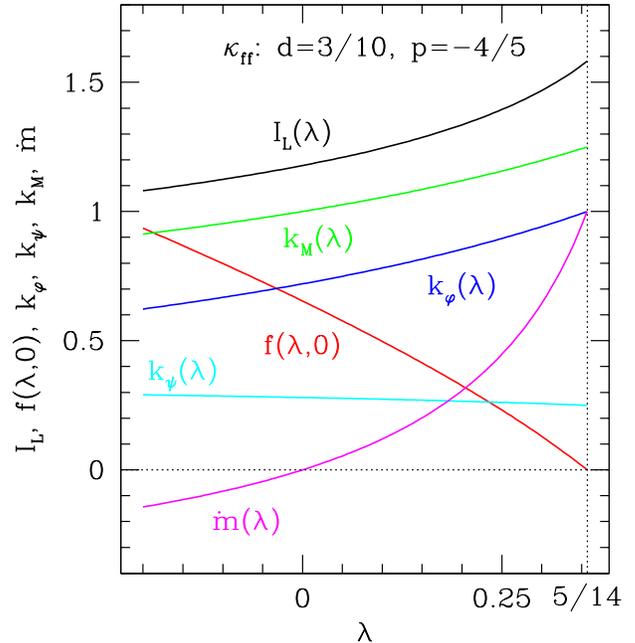}
\caption{Same as Figure \ref{fig:con_es} but for a disk with the free-free opacity ($d=3/10$, $p=-4/5$). \label{fig:con_ff}}
\end{figure}

Comparing Figures \ref{fig:con_lin}-\ref{fig:con_ff} one can notice several features of our new solutions common to both the linear (\S \ref{sect:lin}) and the nonlinear cases. First, decreasing $\lambda$ and $f^\prime_\xi(0)$ always results in the monotonic increase of $f(0)$. This is expected since more severe suppression of accretion (lower $\lambda$) requires stronger central torque. Central mass outflow like in Be stars requires even higher levels of the angular momentum injection by the central object. A unique relation between $\lambda$, which characterizes the efficiency of accretion (see \S \ref{sect:acc_supp}), and $f(0)$, which sets the central torque, is one of the most important properties of the new self-similar solutions.

Second, the $\xi$-extent of our solutions slowly decreases as $\lambda$ is lowered. This is a consequence of our constraint $I_M=1$ for all $\lambda$, which forces higher amplitude solutions to occupy smaller interval of $\xi$.

%%%%%%%%%%%%%%%%%%%%%%%%%%%
%%%%%%%%%%%%%%%%%%%%%%%%%%%

%%%%%%%%%%%%%%%%%%%%%%%%%%%
%%%%%%%%%%%%%%%%%%%%%%%%%%%

%%%%%%%%%%%%%%%%%%%%%%%%%%%
%%%%%%%%%%%%%%%%%%%%%%%%%%%

%%%%%%%%%%%%%%%%%%%%%%%%%%%%%%%%%%%%%%%%%%%%%%%%%%%%%%%%%%%

\subsection{Suppression of accretion}  
\label{sect:acc_supp}

%%%%%%%%%%%%%%%%%%%%%%%%%%%%%%%%%%%%%%%%%%%%%%%%%%%%%%%%%%%

Central torque acting on the disk suppresses mass accretion onto the central object or even reverses it to an outflow. We quantify the degree of this suppression via a dimensionless parameter $\dot m(\lambda)$ defined as 
\ba
\dot m(\lambda) = \frac{\dot M(\lambda,M_d,L_d)}{\dot M(\lambda_0,M_d,L_d)}.
\label{eq:dotm_def}
\ea   
Here $\dot M(\lambda,M_d,L_d)$ is the central accretion rate for a solution corresponding to a disk with total mass $M_d$, angular momentum $L_d$, and a given value of $\lambda$; $\dot M(\lambda_0,M_d,L_d)$ is the value of that rate for a solution with no central torque (see \S \ref{sect:zero_torque}), when $\lambda=\lambda_0$ (see equation (\ref{eq:lamLd})) and central $\dot M$ attains its {\it maximum possible} value for the fixed $M_d$ and $L_d$.

Using equations (\ref{eq:M1})-(\ref{eq:dotMxi}), (\ref{eq:rels}), and keeping in mind our  constraint $I_M=1$, one can easily show that 
\ba   
\dot M(\lambda,M_d,L_d)=\lambda\left[D_{J,0}M_d\left(\frac{I_L(\lambda)M_d}{L_d}\right)^{2-p-d}\right]^{1/(1-d)}.
\label{eq:Mdot_expr}
\ea    
Thus, for a given $\lambda$ the central mass accretion rate is uniquely set by the total disk mass $M_d$ and angular momentum $L_d$. 

Plugging this into the definition (\ref{eq:dotm_def}) we immediately find that 
\ba
\dot m(\lambda) = (2-p)\lambda\left[\frac{I_L(\lambda)}{I_L(\lambda_0)}\right]^{(2-p-d)/(1-d)},
\label{eq:dotm}
\ea   
independent of $M_d$ and $L_d$ and with $I_L(\lambda_0)$ given by equation (\ref{eq:IL_lin}) in the linear and (\ref{eq:ILlam0}) in the nonlinear case. By construction, it is always true that $\dot m(\lambda_0)=1$, i.e. mass inflow is {\it completely unsuppressed} for a disk with zero central torque ($\lambda=\lambda_0$). It is also clear that $\dot m=0$ for a disk with $\lambda=0$, which has {\it fully suppressed accretion}, $\dot M(l=0)=0$. 

\begin{figure}
\centering
\includegraphics[width=0.5\textwidth]{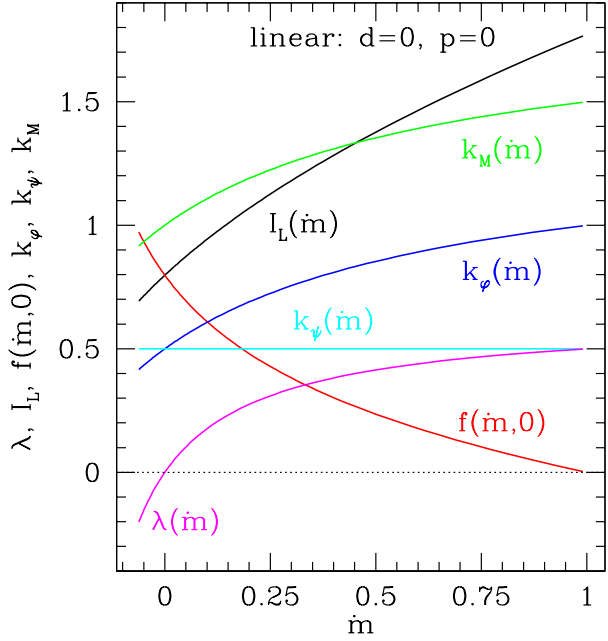}
\caption{Similar to Figure \ref{fig:con_lin} but now the characteristics of our solutions are shown as functions of $\dot m$ --- the degree to which central accretion is suppressed by the central torque. The behavior of $\lambda(\dot m)$ is shown as well. 
\label{fig:con_mdot_lin}}
\end{figure}

\begin{figure}
\centering
\includegraphics[width=0.5\textwidth]{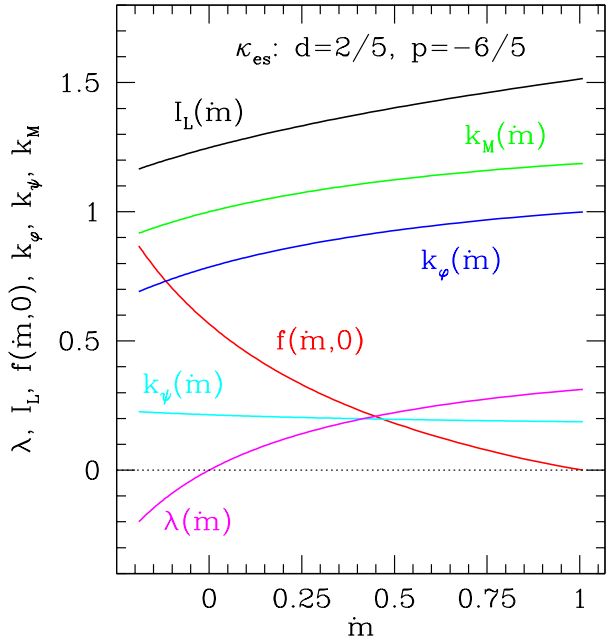}
\caption{Same as Figure \ref{fig:con_mdot_lin} but for a disk with the electron scattering opacity ($d=2/5$, $p=-6/5$). 
\label{fig:con_mdot_es}}
\end{figure}

\begin{figure}
\centering
\includegraphics[width=0.5\textwidth]{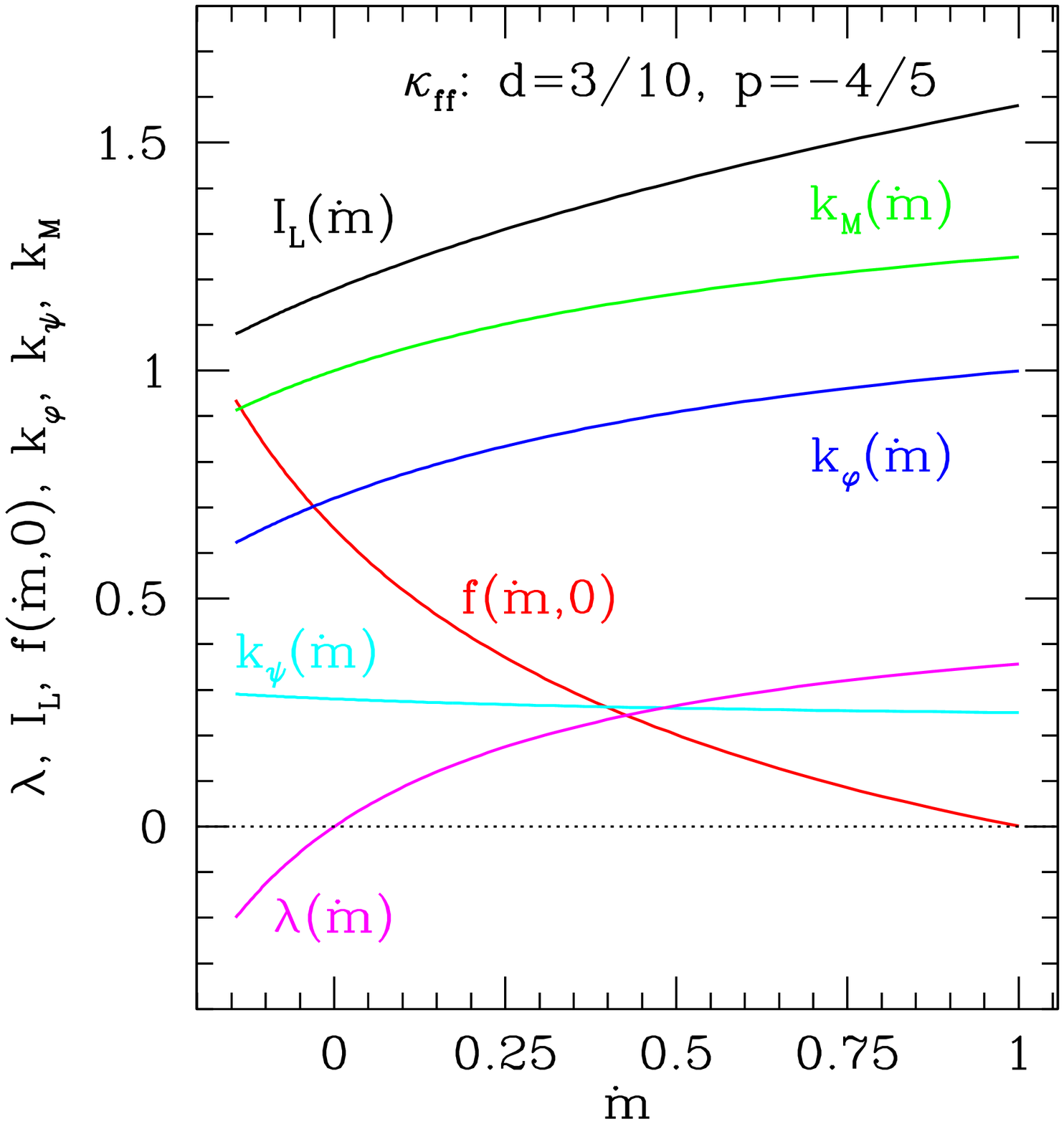}
\caption{Same as Figure \ref{fig:con_mdot_lin} but for a disk with the free-free opacity ($d=3/10$, $p=-4/5$). 
\label{fig:con_mdot_ff}}
\end{figure}

Thus, $\dot m$ indeed represents a convenient dimensionless variable for characterizing the degree of the central inflow suppression for a solution with a given $\lambda$, which is completely independent of the current disk characteristics ($M_d$ and $L_d$). It monotonically varies from $1$ to 0 as we move away from the zero torque solution (\ref{eq:solLdconst}) towards the zero inflow solution (\ref{sect:zero_inflow}), and then  becomes negative for outflow solutions ($\lambda<0$), see Figure \ref{fig:scheme}. Given this, it is useful to show different properties of our solutions as functions of $\dot m$ rather than $\lambda$ (see \S \ref{sect:sol_sup} for a discussion of this issue), and this is done in Figures \ref{fig:con_mdot_lin}-\ref{fig:con_mdot_ff}. 

%%%%%%%%%%%%%%%%%%%%%%%%%%%
%%%%%%%%%%%%%%%%%%%%%%%%%%%

%%%%%%%%%%%%%%%%%%%%%%%%%%%
%%%%%%%%%%%%%%%%%%%%%%%%%%%

%%%%%%%%%%%%%%%%%%%%%%%%%%%
%%%%%%%%%%%%%%%%%%%%%%%%%%%

%%%%%%%%%%%%%%%%%%%%%%%%%%%%%%%%%%%%%%%%%%%%%%%%%%%%%%%%%%%

\subsection{Time evolution}  
\label{sect:time_evolve}

%%%%%%%%%%%%%%%%%%%%%%%%%%%%%%%%%%%%%%%%%%%%%%%%%%%%%%%%%%%

We now go back to equation (\ref{eq:time_eq}) and determine the time evolution of our solutions by solving it:
\ba
\varphi(t)&=&\left[1+\frac{k_\varphi^{-1}}{1-d}\frac{t}{t_0}\right]^{-k_\varphi},
\label{eq:time_ev}\\
%%%%%%%%%%%%%%
k_\varphi &\equiv & \left[d+\frac{(1-\lambda)(1-d)(2-p)}{1-p}\right]^{-1}.
\label{eq:k_varphi}
\ea
We set $\varphi(0)=1$ since we have freedom of choosing this value due to its degeneracy with $F_0$. The scaling exponent $k_\varphi=1$ for $\lambda=(2-p)^{-1}$, i.e. for a disk with zero torque at the center. It monotonically decreases for lower $\lambda$, and $k_\varphi\to 0$ as $\lambda\to -\infty$. This behavior is illustrated in Figures \ref{fig:con_lin}-\ref{fig:con_ff}. By considering $\lambda$ as a free parameter we naturally recover the existing results for $\lambda=\lambda_0$ and $\lambda=0$ \citep{Pringle}.

%%%%%%%%%%%%%%%%%%%%%%%%%%%
\begin{figure*}
\centering
\includegraphics[width=1\textwidth]{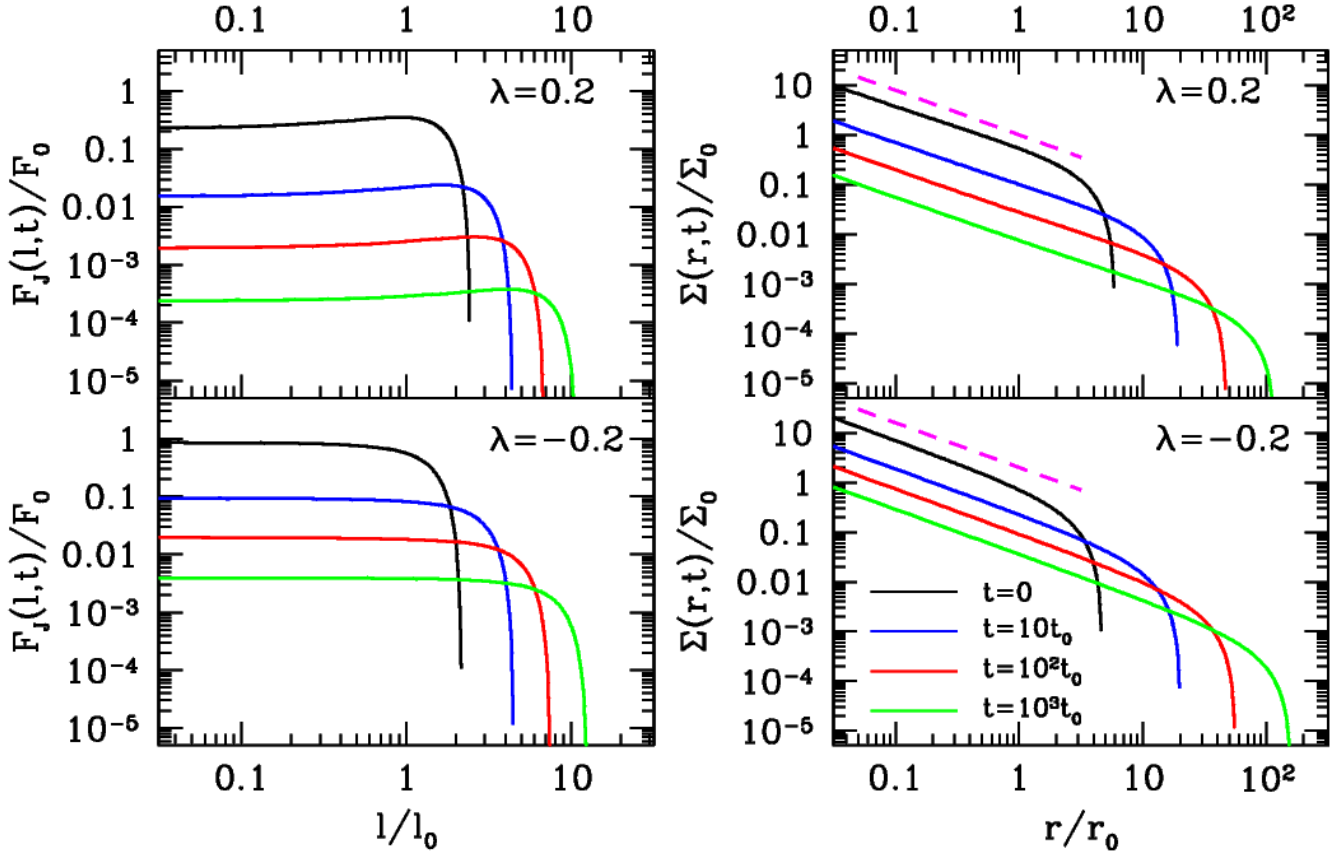}
\caption{Illustration of the self-similar viscous spreading of a disk with the electron scattering opacity ($d=2/5$, $p=-6/5$) for two values of $\lambda$: (top) $\lambda=0.2$ corresponding to partial (by $58\%$) suppression of accretion by the central torque, and (bottom) $\lambda=-0.2$ corresponding to mass injection by the central source. Profiles of $\FJ(l)$ and $\Sigma(r)$ (its normalization is arbitrary) are shown at four different moments of time (ranging from $t=0$ to $t=10^3t_0$) as labeled on the panels; characteristic radial scale $r_0=l_0^2/(GM_c)$. Dashed lines show the universal behavior of the surface density $\Sigma\propto r^{-(p+3)/2}$ near the origin.
\label{fig:self_sim}}
\end{figure*}
%%%%%%%%%%%%%%%%%%%%%%%%%%%

Since $k_\varphi>0$ one can see that $\varphi\to\infty$ at a finite time $t=-(1-d)k_\varphi t_0$ in the past, reflecting our similarity assumption.

Spatial scale of the mass distribution in the disk is set by the condition $\xi\sim 1$, corresponding to a characteristic value of $l=l_0\psi(t)$, varying in time as $l=l_0 \left[1+k_\varphi^{-1}(1-d)^{-1}t/t_0\right]^{k_\psi}$ with $k_\psi=-\delta k_\varphi$, or
\ba
k_\psi \equiv \left[2-p+\frac{d(1-p)}{(1-\lambda)(1-d)}\right]^{-1},
\label{eq:k_psi}
\ea
where we used definition (\ref{eq:lambda}) to eliminate $\delta$. Figures \ref{fig:con_lin}-\ref{fig:con_mdot_ff} demonstrate rather weak dependence of $k_\psi$ on $\lambda$ or $\dot m$.

Plugging the result (\ref{eq:time_ev}) into equations (\ref{eq:ss_ansatz}), (\ref{eq:M1}), (\ref{eq:L1}), (\ref{eq:dotMxi}) we can determine the time evolution of other disk characteristics. In particular, for $t\gg t_0$ the central torque scales as
\ba    
\FJ(0,t)\propto f(0)t^{-k_\varphi}.
\label{eq:torque_time}
\ea    
In the same limit the central inflow rate behaves as
\ba   
\dot M(t)& \propto &\lambda t^{-k_M},
\label{eq:Mdot_ev}\\
%%%%%%%%%%%%
k_M &\equiv & k_\varphi\left[1+\frac{(1-\lambda)(1-d)}{1-p}\right].
\label{eq:k_M}
\ea   
where we used equation (\ref{eq:rels}). Figures \ref{fig:con_lin}-\ref{fig:con_ff} show that $k_M$ monotonically decreases from $k_M=(2-d-p)/(2-p)$ for $\lambda=\lambda_0$ to $k_M\to (2-p)^{-1}$ as $\lambda\to -\infty$. 

The total disk mass and angular momentum vary as 
\ba    
M_d(t) &=& M_d(0)\left[\varphi(t)\right]^{\lambda(1-d)}\propto t^{-\lambda k_\varphi (1-d)},
\label{eq:Md}\\
%%%%%%%%%%%
L_d(t)&=& L_d(0)\left[\varphi(t)\right]^{1-k_\varphi^{-1}}\propto t^{1-k_\varphi},
\label{eq:Ld}
\ea    
where the explicit scalings with time pertain to $t\gg t_0$. 

According to these scalings, a disk with $\lambda=\lambda_0$ has constant $L_d$ (since $k_\varphi\to 1$). Using the conversion in Appendix \ref{sect:conversion} one can easily show that the behavior (\ref{eq:Mdot_ev})-(\ref{eq:k_M}) in such a disk coincides with the one obtained by \citet{Pringle} in the case of a vanishing central torque. Analogously, a disk with $\lambda=0$ preserves its total mass, while the scaling (\ref{eq:torque_time}) of its central torque agrees with the Pringle's result for the case of zero mass inflow.

More generally, it is clear that the disk angular momentum can only increase with time, irrespective of $\lambda>\lambda_0$. Similarly, central torque always decreases. At the same time, disk mass grows for $\lambda<0$ (decretion) and decays for $\lambda>0$ (accretion). Thus, depending on the value of $\lambda$ there could be four possible initial states for the self-similar disk evolution at $t=-(1-d)k_\varphi t_0$:
\begin{enumerate}
    
    \item $\lambda=\lambda_0$: infinite $M_d(0)$, and finite $L_d(0)$ that stays constant through the evolution,
    
    \item $0<\lambda<\lambda_0$: infinite $M_d(0)$, and $L_d(0)=0$,
      
    \item $\lambda=0$: finite $M_d(0)$ that remains fixed through the evolution, and $L_d(0)=0$,
      
    \item $\lambda<0$: $M_d(0)=0$ and $L_d(0)=0$.
  
\end{enumerate}

In Figure \ref{fig:self_sim} we illustrate the self-similar time evolution of a couple of solutions for a disk with the electron scattering opacity. One is an accreting solution with $\lambda=0.2$, corresponding to a $58\%$ suppression of central $\dot M$ compared to the zero-torque case, i.e. $\dot m=0.42$, see Figure \ref{fig:con_es}. Another describes a decretion disk solution with $\lambda=-0.2$, for which $\dot m=-0.19$.

Quite naturally, one finds that near the origin $\FJ(l)$ develops a flat profile (as long as $\lambda>\lambda_0$), the amplitude of which goes down with time. In addition, the radial profile of the surface density $\Sigma$ attains a universal slope for $l\ll l_0\psi(t)$. This naturally follows  from equation (\ref{eq:Sig}), which predicts that 
\ba    
\Sigma(r)\propto r^{-(p+3)/2},~~~~r\to 0,
\label{eq:Sig_slope}
\ea    
since $\FJ(r)\to$ const when there is a non-zero torque at the disk center. Note that this universal result holds for both linear ($d=0$) and nonlinear ($d\neq 0$) settings, and for systems with both accretion ($\lambda>0$) and ejection ($\lambda<0$) of mass by the central object.

Over time the distributions of both $\FJ$ and $\Sigma$ extend over larger and larger distance $r$ (or $l$), while at the same time going down in amplitude. Note that at a given radius $\Sigma$ decreases noticeably slower for the solution with $\lambda<0$, simply because it represents a decretion disk gaining mass at the center. Since $d\neq 0$ for $\kappa=\kappa_{\rm es}$, these solutions are always truncated at finite radii.

%%%%%%%%%%%%%%%%%%%%%%%%%%%
\begin{figure}
\centering
\includegraphics[width=0.5\textwidth]{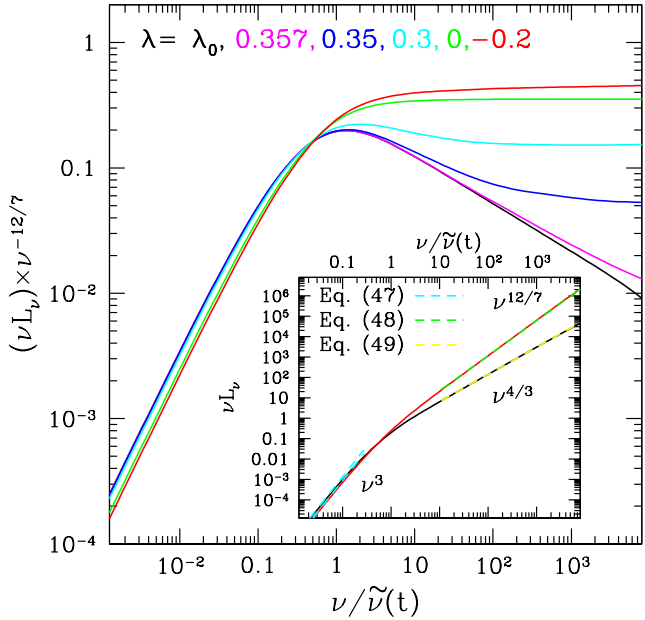}
\caption{Shape of the decretion disk SED for different values of $\lambda$ (colors of different curves correspond to colors of $\lambda$ labels). A self-luminous disk with $\kappa=\kappa_{\rm ff}$ ($d=3/10$, $p=-4/5$) is assumed. The main panel shows $\nu{\cal L}_\nu$ divided by $(\nu/\tilde\nu)^{12/7}$ to simplify comparison for different $\lambda$. Inset shows $\nu{\cal L}_\nu$ for $\lambda=\lambda_0=5/14\approx 0.35714$ and $\lambda=-0.2$ (ejection solution) only, together with the dashed lines illustrating different asymptotic behaviors given by equations in the text, as shown in the inset. Note a significant difference in SED behavior at high frequencies for the disk with no central torque ($\lambda=\lambda_0$, asymptotically scaling as $\nu^{4/3}$) and other values of the similarity parameter (asymptotically converging to $\propto \nu^{12/7}$).
\label{fig_SED_shape}}
\end{figure}
%%%%%%%%%%%%%%%%%%%%%%%%%%%

%%%%%%%%%%%%%%%%%%%%%%%%%%%%%%%%%%%%%%%%%%%%%%%%%%%%%%%%%%%
%%%%%%%%%%%%%%%%%%%%%%%%%%%%%%%%%%%%%%%%%%%%%%%%%%%%%%%%%%%

\section{Observational signatures}
\label{sect:obs}

%%%%%%%%%%%%%%%%%%%%%%%%%%%%%%%%%%%%%%%%%%%%%%%%%%%%%%%%%%%

Viscous stresses driving the outward expansion of the disk inevitably result in energy dissipation, heating the disk and giving rise to observable signatures. The thermal state of the decretion disks is often determined by some  external agents, e.g. via the irradiation by the central object. This is expected to be the case in the Be disks \citep{Bes} or in the outer parts of the protoplanetary disks around young stellar binaries \citep{Vartanyan}. In this regime internal dissipation is not expected to appreciably affect the disk spectrum.

However, in many systems thermodynamics of the disk is dominated by the internal heating. This is likely to be true for the disks orbiting compact objects (e.g. neutron stars in the propeller regime), inner regions of the circumbinary protoplanetary disks \citep{Vartanyan}, circumbinary disks around supermassive black hole binaries \citep{Rafikov2013}, etc. In these systems the details of the internal dissipation get directly reflected in their spectral energy distribution (SED), thus providing an observational probe of the disk physics. For that reason in this section we will focus on describing the SEDs of such {\it self-luminous} decretion disks, heated primarily by the viscous dissipation.

In Appendix \ref{sect:SED} we show that the SED of a self-luminous self-similar decretion disk is given by 
\ba     
\nu {\cal L}_\nu(t) = F_0\Omega_0\left[\varphi(t)\right]^{1-3\delta}\Phi\left(\lambda,\frac{\nu}{\tilde\nu(t)}\right),
\label{eq:SED}
\ea
($\Omega_0=(GM_c)^2 l_0^{-3}$) where the frequency dependence of the spectrum is characterized by the shape function
\ba 
\Phi(\lambda,z) \equiv \frac{45}{\pi^4}z^4\int\limits_0^\infty\frac{\xi^3d\xi}{\exp\left[z\left(f(\lambda,\xi)\xi^{-7}\right)^{-1/4}\right]-1}.
\label{eq:Phi}
\ea
The characteristic frequency 
\ba
\tilde\nu(t) &\equiv &\nu_0\left[\varphi(t)\right]^{(1-7\delta)/4},
\label{eq:tildenu}\\
%%%%%
\nu_0 &\equiv &\frac{k_B}{h}\left[\frac{3}{8\pi}\frac{(GM)^4F_0}{\sigma l_0^7}\right]^{1/4},
\label{eq:nu0}
\ea     
monotonically decreases with time as $\delta<0$, see equations (\ref{eq:lambda}) and (\ref{eq:time_ev}). 

It is obvious that this SED exhibits a self-similar behavior anchored to the evolution of the characteristic frequency $\tilde\nu$. The frequency dependence of the SED is uniquely determined by the dependence of $\Phi(\lambda,z)$ on $z$, which is illustrated in Figure \ref{fig_SED_shape} for several values of $\lambda$. 

%%%%%%%%%%%%%%%%%%%%%%%%%%%
\begin{figure}
\centering
\includegraphics[width=0.5\textwidth]{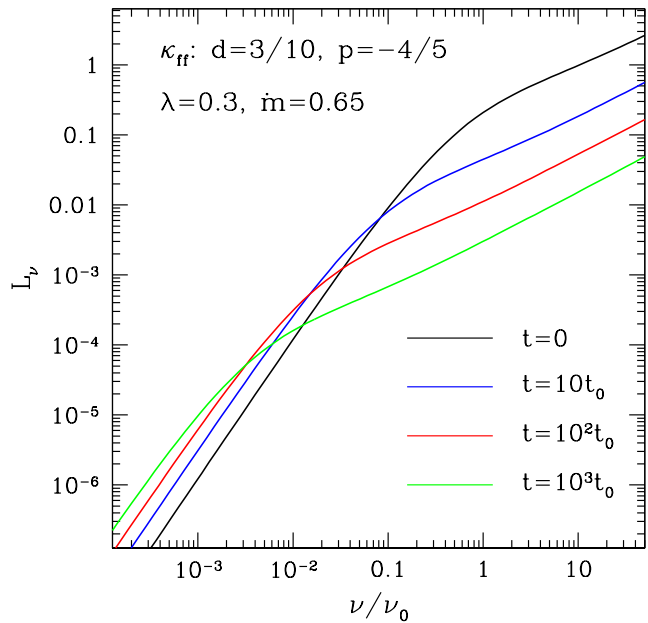}
\caption{Evolution of the disk spectrum (here shown as ${\cal L}_\nu$) as a function of time. Calculation is done for the disk parameters used in Figure \ref{fig_SED_shape} and assumes $\lambda=0.3$ (accretion suppression $\dot m\approx 0.65$). Different curves correspond to different times as labeled on the Figure.
\label{fig:SED_evolve}}
\end{figure}
%%%%%%%%%%%%%%%%%%%%%%%%%%%

One can see that in the low frequency limit $\nu\ll \tilde\nu$ the scaling of the shape function $\Phi$ with $\nu$ is universal and independent of $\lambda$. This can be easily understood from the asymptotic behavior of $\Phi$, which is discussed in Appendix \ref{sect:SED}. For $\nu\ll \tilde\nu(t)$ one finds
\ba    
\Phi(\lambda,z)\to c_6(\lambda)z^3,~~~z\ll 1,
\label{eq:low_freq}
\ea  
with $c_6$ given by equations (\ref{eq:c6}). This behavior describes the Raileigh-Jeans tail of the disk emission and is robust for all $\lambda$, as Figure \ref{fig_SED_shape} shows. The $\lambda$-dependence of  the amplitude of this asymptotic is also rather weak. 

Things are different in the high frequency limit, $\nu\gg\tilde\nu(t)$. There one finds, as long as $\lambda\neq \lambda_0$, that 
\ba    
\Phi(\lambda,z)\to c_7\left[f(\lambda,0)\right]^{4/7}z^{12/7}, ~~~z\gg 1,
\label{eq:high_freq}
\ea     
with $c_7$ given by equation (\ref{eq:c7}). Because of this scaling we chose to divide $\nu{\cal L}_\nu$ by $\nu^{12/7}$ in Figure \ref{fig_SED_shape} to better illustrate the SED dependence on $\lambda$ (the inset of that Figure shows SED for two values of $\lambda$ without such division).

However, for the disk without any torque at the center ($\lambda=\lambda_0$ and $f(\lambda_0,0)=0$) the SED behavior is qualitatively different:
\ba   
\Phi(\lambda_0,z)\to c_8 (2-p)^{-2/3}z^{4/3}, ~~~z\gg 1,
\label{eq:high_freq_no_torque}
\ea     
with $c_8$ given by equation (\ref{eq:c8}). Asymptotic behaviors (\ref{eq:low_freq})-(\ref{eq:high_freq_no_torque}) are illustrated in the inset in Figure \ref{fig_SED_shape}.

The difference in the high-$\nu$ SED scaling between the standard accretion disk (\ref{eq:high_freq_no_torque}) and the disk with some nonzero central torque (\ref{eq:high_freq}) was previously noted in \citet{Syer} and \citet{Rafikov2013}. Figure \ref{fig_SED_shape} clearly shows that as long as $\lambda$ even slightly deviates from $\lambda_0$, the high-frequency asymptotic of $\Phi$ follows the behavior (\ref{eq:high_freq}). Already at $\lambda=0.35$ (different from $\lambda_0$ by only $0.07$) $\Phi(\lambda,z)$ clearly tends to converge to $\nu^{12/7}$ scaling. At smaller values of $\lambda$, including the negative ones, the convergence is faster and only the amplitude of the scaling depends on $\lambda$.

Temporal evolution of the SED accompanying the viscous spreading of the disk is illustrated in Figure \ref{fig:SED_evolve}. There we show $\nu {\cal L}_\nu$ at several moments of time for a self-luminous decretion disk with $\kappa=\kappa_{\rm ff}$ ($d=3/10$, $p=-4/5$) and $\lambda=0.3$ (for which accretion is suppressed by $35\%$, i.e. $\dot m\approx 0.65$, see Figure \ref{fig:con_lin}). Over time, the spectrum of the disk shifts towards lower frequencies, while maintaining its overall self-similar shape.

One can see that as time goes by the spectral power above the characteristic frequency $\tilde \nu$ always decreases. Indeed, using equation (\ref{eq:high_freq}) one finds $\nu {\cal L}_\nu(t)\propto (\nu/\nu_0)^{12/7}\left[\varphi(t)\right]^{4/7}$ for $\nu\gg \tilde\nu(t)$, meaning the {\it decay} of ${\cal L}_\nu$ for $\nu\gtrsim \tilde\nu$.

On the contrary, below $\tilde \nu$ the amplitude of ${\cal L}_\nu$ {\it grows} with time. This can be understood by combining equations (\ref{eq:SED})-(\ref{eq:tildenu}) and (\ref{eq:low_freq}) to find that $\nu {\cal L}_\nu(t)\propto (\nu/\nu_0)^3\left[\varphi(t)\right]^{(1+9\delta)/4}$ for $\nu\ll\tilde \nu$. As $\delta<0$, this implies that $\nu {\cal L}_\nu$ increases with time at a fixed frequency, as long as $\nu$ stays below $\tilde\nu(t)$. 

From equation (\ref{eq:dEdr}) it is easy to see that the bolometric luminosity ${\cal L}$ of a decretion disk is proportional to $\FJ(r\to 0)$, as long as the latter is non-zero (to obtain ${\cal L}$ the integral of $d\dot E_v/dr$ has to be truncated at some inner radius). As a result, ${\cal L}\propto\varphi(t)\propto t^{-k_\varphi}$ for such disks. This luminosity evolution is different from that of the disks with zero central torque ($\lambda=\lambda_0$), which have ${\cal L}\propto\dot M(t)\propto t^{-k_M}$, see equation (\ref{eq:Mdot_ev}).

The difference between the high-frequency spectra given by equations (\ref{eq:high_freq}) and (\ref{eq:high_freq_no_torque}) can be used as an observational probe of the presence of the non-zero torque at the center of a decretion disk. \citet{Rafikov2013} suggested utilizing this feature as a way of inferring the presence of the binary supermassive black holes from the quasar spectra. Based on Figure \ref{fig_SED_shape} we expect ${\cal L}_\nu$ to have a  steeper $\nu$-dependence (\ref{eq:high_freq}) at high frequencies as long as there is even weak central torque at the center. Spectrum of a disk with no central torque whatsoever would follow the shallower frequency dependence (\ref{eq:high_freq_no_torque}). However, in has to be remembered that this distinction applies only to a purely self-luminous disk radiating as a black body. Any deviation from this regime (e.g. illumination by the central object, strong emission lines, etc.) could easily affect this observational probe of the central torque.

%%%%%%%%%%%%%%%%%%%%%%%%%%%%%%%%%%%%%%%%%%%%%%%%%%%%%%%%%%%
%%%%%%%%%%%%%%%%%%%%%%%%%%%%%%%%%%%%%%%%%%%%%%%%%%%%%%%%%%%

\section{Discussion}
\label{sect:disc}

%%%%%%%%%%%%%%%%%%%%%%%%%%%%%%%%%%%%%%%%%%%%%%%%%%%%%%%%%%%

Our results in \S \ref{sect:new} clearly show that to fully specify the self-similar disk evolution one must provide the values of the three constants --- $\lambda$, $F_0$, and $l_0$. There are different ways in which they can be fixed by the physics of the problem at hand, which we discuss next in \S \ref{sect:sol_sup}-\ref{sect:sol_torque}. 

Once this is done, one obtains a lot of information about the integral characteristics of the decretion disk evolution, for a given value of $\lambda$. In particular, one finds a unique relation between the central $\dot M(0,t)$ and torque $F_J(0,t)$ (Figures \ref{fig:con_lin}-\ref{fig:con_ff}), determines time evolution of the total angular momentum $L_d$ and mass $M_d$ of an evolving disk, the rate at which it expands, and so on (\S \ref{sect:time_evolve}). We illustrate the use of these results in Rafikov (2016, in preparation), where we employ our understanding of the decretion disk evolution to constrain physical mechanisms of the eccentricity excitation in the post-main sequence binaries.

%%%%%%%%%%%%%%%%%%%%%%%%%%%%%%%%%%%%%%%%%%%%%%%%%%%%%%%%%%%

\subsection{Solution determination: fixed degree of the suppression of accretion}  
\label{sect:sol_sup}

%%%%%%%%%%%%%%%%%%%%%%%%%%%%%%%%%%%%%%%%%%%%%%%%%%%%%%%%%%%

We now discuss how one can uniquely determine $\lambda$, $F_0$ and $l_0$ using physical arguments relevant for different astrophysical objects. 

The value of $\lambda$ can be fixed if one expects the torque exerted by the central object to suppress $\dot M$ at the origin by a {\it prescribed} amount $\dot m$ compared to the accretion rate in the absence of the central torque. A direct and monotonic relation between $\lambda$ and $\dot m$ established in \S \ref{sect:acc_supp} then allows one to determine the former once the latter is fixed. In particular, in the linear case one would invert analytical formula (\ref{eq:dotm_lin}) for that purpose, using the definition (\ref{eq:ab}). In the nonlinear case ($d\neq 0$) one would use the numerical calculations such as described in \S \ref{sect:acc_supp} and shown in Figures \ref{fig:con_mdot_lin}-\ref{fig:con_mdot_ff}. 

Once $\lambda$ is fixed, the values of $F_0$ and $l_0$ are  uniquely specified by the total mass $M_d(0)$ and angular momentum $L_d(0)$ of the disk at $t=0$. This is shown mathematically via equations (\ref{eq:F0})-(\ref{eq:l0}) in Appendix \ref{sect:sol_det}. Thus, the knowledge of $M_d$ and $L_d$ at some moment in time (which can always be set to $t=0$) fully specifies the subsequent self-similar evolution of the disk. 

Note that there are other ways of fixing $F_0$ and $l_0$ for a given $\lambda$. For example, instead of $L_d(0)$ one may choose to specify the characteristic radius enclosing a given fraction of the disk mass --- obviously, it is directly related to $l_0$. There are many other similar choices, which we do not discuss here. 

Fixing the value of $\lambda$ via known $\dot m$ is a very simple and attractive way of specifying the disk evolution. For example, recent simulations of the circumbinary disks \citep{MacFadyen2008,DOrazio2013,Farris2014} provide a measurement of the accretion rate in the presence of the binary torque, resulting in the estimate of $\dot m$ and, thus, $\lambda$. Motivated by these results, \citet{Martin} explored one-dimensional viscous evolution of the circumbinary disks including a model with a fixed non-zero value of $\dot m$, similar to what we have decsribed.

%%%%%%%%%%%%%%%%%%%%%%%%%%%%%%%%%%%%%%%%%%%%%%%%%%%%%%%%%%%

\subsection{Solution determination: known physics of the central barrier}  
\label{sect:sol_barrier}

%%%%%%%%%%%%%%%%%%%%%%%%%%%%%%%%%%%%%%%%%%%%%%%%%%%%%%%%%%%

The problem with the approach outlined in \S \ref{sect:sol_sup} is that there is usually no a priori reason why one should expect $\dot m$ to be constant in time. Indeed, the central $\dot M$ is set by the physics of the central barrier to accretion (details of the torque exerted on the disk by the accreting object, local gas density, etc.), while $\dot M(\lambda=\lambda_0)$ is set by the {\it global} structure of the disk.

At the same time, it is still possible to find a unique value of the similarity parameter $\lambda$ using the knowledge of what sets $\dot M$ at the disk center. It is reasonable to expect that $\dot M$ should be proportional to the amount of mass in the inner disk, i.e. to $\Sigma(r\to 0)$. Since $\Sigma$ is related to $\FJ$ via the definition (\ref{eq:F_J}) we will consider a simple boundary condition  for $\dot M$ in the power law form
\ba    
\dot M=K\left[F_J(r\to 0)\right]^\eta,
\label{eq:prescr1}
\ea    
with constant $K$ and $\eta>0$. In other words, the larger is the inner torque $F_J(r\to 0)$, the more mass accumulates near the origin, the higher is $\Sigma$ there, and the larger is $\dot M$.

Using equations (\ref{eq:ss_ansatz}), (\ref{eq:dotMxi}) and (\ref{eq:lambda}) one can see that with this prescription for the physics of the inner barrier the self-similarity uniquely determines
\ba   
\lambda=1+(1-\eta)\frac{1-p}{1-d}.
\label{eq:lam_phys}
\ea   

The same procedure also yields an algebraic relation between $F_0$ and $l_0$. To separately determine their values one needs to supply additional information such as done in \S \ref{sect:sol_sup}. We will assume here that we know disk mass $M_d$ at time $t=0$. Then, using equation (\ref{eq:M1}) with $\varphi(0)=I_M=1$, one finds that $F_0$ and $l_0$ are given by equations (\ref{eq:F0_phys}) and (\ref{eq:l0_phys}). 
Given the constraint (\ref{eq:phys_meaning}) our derived value (\ref{eq:lam_phys}) of $\lambda$ implies 
\ba    
\eta\ge 1+\frac{1-d}{2-p},
\label{eq:eta_constr}
\ea    
i.e. that self-similarity is possible only for {\it steep enough} dependence of the central $\dot M$ on $F_J(r\to 0)$, certainly faster than linear. Thus, the inner barrier should be less effective at suppressing gas inflow as more gas accumulates near the origin.

%%%%%%%%%%%%%%%%%%%%%%%%%%%%%%%%%%%%%%%%%%%%%%%%%%%%%%%%%%%

\subsection{Solution determination: prescribed central $\dot M$ or torque $\FJ$}  
\label{sect:sol_torque}

%%%%%%%%%%%%%%%%%%%%%%%%%%%%%%%%%%%%%%%%%%%%%%%%%%%%%%%%%%%

It is also possible that the system imposes boundary conditions on the disk that enforce self-similarity of its evolution. For example, consider a central object that {\it ejects} mass at a rate which asymptotically scales as a power law of time, $\dot M\propto t^{-\eta_M}$, where $\eta_M$ is a constant determined by the physics of the ejection process. An example of such system could be a Be star or a post-main sequence binary losing mass via its outer Lagrange point. Then equation (\ref{eq:Mdot_ev}) implies that $k_M=\eta_M$, which, acccording to equations (\ref{eq:k_varphi}) and (\ref{eq:k_M}), uniquely determines the value of the similarity parameter $\lambda$ as a function of $\eta_M$ (see the discussion in \S \ref{sect:compare}).

This, in turn, fixes the value of $k_\varphi$ (Eq. [\ref{eq:k_varphi}]) and, according to equation (\ref{eq:torque_time}), sets the time evolution of the central torque $F_J(0,t)$. Then a natural question to ask is whether one would naturally expect $F_J(0,t)$ to follow this particular unique behavior, as required by the similarity of the solution. At least in some cases the answer is yes. 

For example, a stellar binary losing mass through its L2 Lagrange point exerts gravitational torque on the escaping gas at particular resonant locations in the disk \citep{GT80}. The amplitude of this torque is proportional to the disk surface density at the resonant radii and should naturally self-regulate to follow the behavior (\ref{eq:torque_time}) in the following way. If $F_J(0,t)$ grows {\it above} the value needed for the self-similar expansion of the disk, the inner disk will absorb excess angular momentum, driving its expansion. This will reduce $\Sigma$ at the resonant locations until $F_J(0,t)$ is brought back in accord with the global viscous evolution of the disk. 

On the contrary, if at any point in time $F_J(0,t)$ becomes {\it lower} than the self-similar value (\ref{eq:torque_time}), the mass will be less readily evacuated from the central object's vicinity, causing gas pileup at the resonant locations and the return of the central torque to the behavior  (\ref{eq:torque_time}). This is how the central torque would self-regulate to ensure the self-similar behavior determined by the exponent $\eta_M$.

Determination of the values of $F_0$ and $l_0$ is possible in this case via the normalization of $\dot M$, which provides an algebraic relation between these variables. Another relation can be obtained e.g. through the knowledge of the total angular momentum of the disk $L_d$ at some moment of time. Then, similar to \S \ref{sect:sol_sup}-\ref{sect:sol_barrier} one would uniquely determine both $F_0$ and $l_0$. We do not show the resulting expressions due to their complexity even though they could be easily derived as just described.

Another possibility for governing the self-similar evolution is via the prescribed central torque on the disk, which may be more typical for {\it accreting} objects (i.e. $\dot M>0$). If $F_J(0,t)\propto t^{-\eta_L}$ with a constant $\eta_L$, then equation (\ref{eq:torque_time}) immediately implies $k_\varphi=\eta_L$, thus fixing the value of $\lambda$ via the equation (\ref{eq:k_varphi}). Provided that the central $\dot M$ self-regulates to obey equation (\ref{eq:Mdot_ev}) the self-similar evolution would again be possible.

%%%%%%%%%%%%%%%%%%%%%%%%%%%%%%%%%%%%%%%%%%%%%%%%%%%%%%%%%%%

\subsection{Comparison with the existing studies}  
\label{sect:compare}

%%%%%%%%%%%%%%%%%%%%%%%%%%%%%%%%%%%%%%%%%%%%%%%%%%%%%%%%%%%

Following the pioneering work of \citet{lynden-bell_1974} a number of authors have explored viscous evolution of the decretion disks in a variety of contexts. Self-similar solutions, which are the focus of our work, were first discussed for the linear problem ($d=0$) in \citet{lynden-bell_1974}, see \S \ref{sect:zero_torque} and \ref{sect:zero_inflow}. Self-similar ansatz for the nonlinear problem was first discussed in \citet{filipov_1984}, but the detailed analysis of this problem had to wait until \citet{lyubarskij_1987} obtained the two known nonlinear solutions without either the central torque (\ref{eq:solLdconst}) or the central inflow (\ref{eq:solMdconst}). These solutions were also discussed in \citet{Filipov88}, \citet{filipov_1988}, \citet{Cannizzo}, and \citet{Pringle}. Self-similar solutions with somewhat different boundary conditions were studied by \citet{Lipunova}.

Our work extends these past studies by also exploring a much more general class of astrophysical systems in which neither the central inflow nor the central torque vanish. Some qualitative discussion of the decretion disk evolution in this case can be found in \citet{Vartanyan}. Also, in his study of the circumbinary disks around the supermassive black hole binaries \citet{Rafikov2013} found self-similar solutions with both $\dot M(r\to 0)\neq 0$ and $\FJ(r\to 0)\neq 0$ for {\it accretion disks externally supplied at a fixed $\dot M$},
generalizing the previous result of \citet{ivanov_1999}, which was limited to $\dot M(r\to 0)=0$. This is a qualitatively different setup compared to the decretion disks studied here, for which no self-similar solutions with such general boundary conditions have been explored until now.

Moreover, our results also apply to systems, in which a decretion disk is fed with mass {\it ejected by the central object}, such as the disks around Be stars. We are not aware of any existing self-similar solutions applicable to disks with central mass injection, which makes our results particularly valuable for understanding disks of Be stars and mass-losing post-main sequence binaries. 

For example, in his study of the Be disks \citet{Okazaki} numerically calculated viscous evolution of an isothermal decretion disk ($c_s$=const) fed at a constant injection rate $\dot M$. In our self-similar ansatz (\ref{eq:DJpow}) such disk would correspond to $d=0$ and $p=1/2$, as follows from equation (\ref{eq:D_J}) for constant $c_s$. However, equations (\ref{eq:k_varphi}), (\ref{eq:Mdot_ev}), (\ref{eq:k_M}) demonstrate that the assumption of time invariant $\dot M$ is incompatible with the self-similarity of the disk evolution: it would require $k_M\to 0$, which is impossible, see the discussion after equation (\ref{eq:k_M}). This expectation agrees with the numerical results of \citet{Okazaki}, which indeed do not exhibit the development of a self-similar profile of the surface density. 

At the same time \citet{Okazaki} found the convergence of $\Sigma(r)$ to $r^{-2}$ profile previously suggested by \citet{Bjorkman}, which is what equation (\ref{eq:Sig_slope}) predicts for $p=1/2$. However, equation (\ref{eq:Sig_slope}) does not require similarity and follows simply from the fact that $\FJ(r)\to$ const as $r\to 0$ (naturally fulfilled for any disk with central mass source) as discussed earlier in \S \ref{sect:time_evolve}.

%%%%%%%%%%%%%%%%%%%%%%%%%%%%%%%%%%%%%%%%%%%%%%%%%%%%%%%%%%%
%%%%%%%%%%%%%%%%%%%%%%%%%%%%%%%%%%%%%%%%%%%%%%%%%%%%%%%%%%%

\section{Summary}  
\label{sect:sum}

%%%%%%%%%%%%%%%%%%%%%%%%%%%%%%%%%%%%%%%%%%%%%%%%%%%%%%%%%%%

Our work provides general understanding of the decretion disk evolution in the late time asymptotic limit, when the viscous stresses drive the disk structure towards the self-similarity. Going beyond the existing studies, we calculate the self-similar viscous evolution of the most general decretion disks that feature {\it both the nonzero accretion (or decretion) rate at the center and the nonzero central torque}. This situation naturally arises in a number of real astrophysical objects --- accreting neutron stars, post-main sequence binaries, disks of Be stars, etc.

The variety of diverse evolutionary pathways of decretion disks, both linear and nonlinear, is shown to be a function of a single similarity parameter $\lambda$. The two previously known similarity solutions \citep{lynden-bell_1974,Pringle} correspond to the {\it two discrete} values of this parameter (see \S \ref{sect:zero_torque} and \ref{sect:zero_inflow}). With our new results we have now covered a {\it continuum} of other possible values of $\lambda$, relevant for both accretion and ejection of mass by the central object. We have also shown that $\lambda$ is closely related to the degree $\dot m$, to which the nonzero central torque suppresses accretion by the central object (\S \ref{sect:acc_supp}). 

Our calculations reveal the intimate connection between the central torque acting on the disk and the central accretion rate, which is closely related to the value of $\lambda$. Once the latter is known, the self-similar ansatz uniquely predicts in a transparent way the time evolution of the main disk properties --- its full mass and angular momentum, radial scale, central torque, and mass accretion rate.  We calculate observational signatures of the self-luminous decretion disks and show that their spectra are different from the SEDs of the conventional accretion disks with zero central torque. This is also true for the evolution of their bolometric luminosity.  

We then discuss a variety of ways, in which the characteristics of our new self-similar solutions --- their amplitude, radial scale, value of $\lambda$ --- can be constrained for different astrophysical objects. Our results should be applicable to understanding the viscous evolution of the decretion disks in various astrophysical settings (Rafikov 2016, in preparation).

\acknowledgements

R.R.R. is an IBM Einstein Fellow at the IAS. Financial support for this study has been provided by NSF via grants AST-1409524,  AST-1515763, NASA via grants 14-ATP14-0059, 15-XRP15-2-0139, and The Ambrose Monell Foundation.

%%%%%%%%%%%%%%%%%%%%%%%%%%%%%%%%%%%%%%%%%%%%%%%%%%%%%%%%%%%
%%%%%%%%%%%%%%%%%%%%%%%%%%%%%%%%%%%%%%%%%%%%%%%%%%%%%%%%%%%

\bibliographystyle{apj}
\bibliography{references}

\appendix 

%%%%%%%%%%%%%%%%%%%%%%%%%%%%%%%%%%%%%%%%%%%%%%%%%%%%%%%%%%%
%%%%%%%%%%%%%%%%%%%%%%%%%%%%%%%%%%%%%%%%%%%%%%%%%%%%%%%%%%%

\section{Connection to the notation of Pringle (1991)}  
\label{sect:conversion}

%%%%%%%%%%%%%%%%%%%%%%%%%%%%%%%%%%%%%%%%%%%%%%%%%%%%%%%%%%%

\citet{Pringle} studied the nonlinear viscous spreading problem assuming viscosity in the form $\nu\propto\Sigma^m r^n$. Using definitions (\ref{eq:F_J}) and (\ref{eq:D_J}) one can show that such scaling implies
\ba   
d=\frac{m}{m+1},~~~p=\frac{2n-3m-2}{m+1}
\label{eq:conversion}
\ea   
in our notation. Also, \citet{Pringle} wrote down the evolution equation not for $\FJ$ but for 
\ba   
S\propto \Sigma R^3\propto \FJ^{1/(m+1)}l^{(3m+2-2n)/(m+1)}
\ea   
as a function of time and $r\propto l^{1/2}$. These relations allow one to convert analytical solutions (\ref{eq:solLdconst}) and (\ref{eq:solMdconst}) into Pringle's notation. Note that in \citet{Pringle} these solutions are not normalized to satisfy $I_M=1$.

%%%%%%%%%%%%%%%%%%%%%%%%%%%%%%%%%%%%%%%%%%%%%%%%%%%%%%%%%%%
%%%%%%%%%%%%%%%%%%%%%%%%%%%%%%%%%%%%%%%%%%%%%%%%%%%%%%%%%%%

\section{Details of the linear solution}  
\label{sect:lin_details}

%%%%%%%%%%%%%%%%%%%%%%%%%%%%%%%%%%%%%%%%%%%%%%%%%%%%%%%%%%%

Change of variables 
\ba
z\equiv - \kappa(\lambda) \xi^{2-p},~~~\kappa(\lambda)=\frac{1-\lambda}{(1-p)(2-p)},
\label{eq:z}
\ea
converts equation (\ref{eq:xi_eq1}) with $d=0$ into Kummer's equation \citep{Abram}
\ba
&& zf^{\prime\prime}_{zz}+(b-z)f^{\prime}_{z}-af=0,
\label{eq:hyper_eq}\\
%%%%%%%%
&& a(\lambda)\equiv\frac{1-p}{(1-\lambda)(2-p)},~~~b\equiv\frac{1-p}{2-p}.
\label{eq:ab}
\ea
Its solutions can be generally expressed via the confluent hypergeometric function. A particular solution that describes the disk with a finite mass \citep{Abram} and satisfies the condition $I_M=1$ is given by equation (\ref{eq:lin_sol}) with 
\ba   
c_5(\lambda)=(2-p)\left(\kappa(\lambda)\right)^b\frac{\Gamma\left(1+b-a(\lambda)\right)}{\Gamma(b)},
\label{eq:c4}
\ea 
where $\Gamma(t)$ is the $\gamma$-function \citep{Grad}.  In a disk with zero inflow at the center ($\lambda=0$) this pre-factor becomes
\ba     
c_2=\left[\frac{2-p}{(1-p)^{1-p}}\right]^{1/(2-p)}\left[\Gamma\left(\frac{1-p}{2-p}\right)\right]^{-1}.
\label{eq:c5}
\ea     

For linear problem the angular momentum integral $I_L$ can be expressed as an analytic function of the disk parameters:
\ba    
I_L(\lambda)=\kappa^{-1/(2-p)}\frac{\Gamma\left(1+b-a(\lambda)\right)\Gamma(2-b)}{\Gamma\left(2-a(\lambda)\right)\Gamma(b)}.
\label{eq:IL_lin}
\ea   
Using equation (\ref{eq:dotm}) we find the degree to which accretion is suppressed for a given $\lambda$ in the linear case as
\ba     
\dot m(\lambda)=\frac{\lambda}{2-p}\left[\frac{\Gamma\left(1+b-a(\lambda)\right)}{\Gamma\left(2-a(\lambda)\right)\Gamma(b)}\right]^{2-p}.
\label{eq:dotm_lin}
\ea

%%%%%%%%%%%%%%%%%%%%%%%%%%%%%%%%%%%%%%%%%%%%%%%%%%%%%%%%%%%
%%%%%%%%%%%%%%%%%%%%%%%%%%%%%%%%%%%%%%%%%%%%%%%%%%%%%%%%%%%

\section{Details of the nonlinear solutions}  
\label{sect:const}

%%%%%%%%%%%%%%%%%%%%%%%%%%%%%%%%%%%%%%%%%%%%%%%%%%%%%%%%%%%

Here we provide expressions for the various constant factors relevant for the nonlinear problem ($d\neq 0$):
\ba
c_1 &=&\frac{d (2-p)^{d-1}}{2-p-d},~~~
\label{eq:c1}\\
%%%%%
c_3 &=& \left[\left(\frac{d}{1-p}\right)^{1-p}\frac{2-p}{B^{2-p}}\right]^{1/(2-p-d)},~~~
\label{eq:c2}\\
%%%%%
c_4 &=& \left[\left(\frac{d}{1-p}\right)^{1-d}\frac{B^d}{2-p}\right]^{(2-p)/(2-p-d)},
\label{eq:c3}
\ea   
where $B = \mbox{B}\left((1-p)/(2-p),d^{-1}\right)$ is the $\beta$-function. Also angular momentum integrals (\ref{eq:L1}) for zero inflow and zero torque cases are given by 
\ba
I_L(0) &=& \frac{d}{2-p}\frac{c_3^{1-d}}{c_4},
\label{eq:IL0}\\
%%%%%%%%%%
I_L(\lambda_0) &=& \frac{(2-p)^{d-1}}{(2-p-d)c_1^{(3-p-d)/(2-p-d)}}
\mbox{B}\left(\frac{3-p-d}{2-p-d},d^{-1}\right)
\label{eq:ILlam0}
\ea
In the linear case equation (\ref{eq:IL_lin}) should be used instead.

%%%%%%%%%%%%%%%%%%%%%%%%%%%%%%%%%%%%%%%%%%%%%%%%%%%%%%%%%%%
%%%%%%%%%%%%%%%%%%%%%%%%%%%%%%%%%%%%%%%%%%%%%%%%%%%%%%%%%%%

\section{SED calculation}  
\label{sect:SED}

%%%%%%%%%%%%%%%%%%%%%%%%%%%%%%%%%%%%%%%%%%%%%%%%%%%%%%%%%%%

To compute the SED of a self-luminous decretion disk we use the following relation between the viscous energy dissipation rate per unit radius $d\dot E_v/dr$, effective temperature of the disk $T_e$, and $\FJ$ \citep{Rafikov2013}:
\ba   
\frac{d\dot E_v}{dr}=4\pi r\sigma T_e^4=\frac{3}{2}\frac{\FJ \Omega}{r},
\label{eq:dEdr}
\ea   
where we assumed Keplerian rotation. Given the self-similar behavior of $\FJ$ in the form (\ref{eq:ss_ansatz}) with the known $\varphi(t)$, $\psi(t)$, and $f(\xi)$, one immediately finds the behavior of $T_e$ as a function of $r$ and $t$.

We compute disk SED as $\nu {\cal L}_\nu=2\pi\nu\int_0^\infty 2\pi r B_\nu(T_e(r,t),\nu)dr$, where $B_\nu$ is a Planck function. After a series of straightforward transformation, we find the SED to be given by the equations (\ref{eq:SED})-(\ref{eq:nu0}), where we used equation (\ref{eq:phi_psi}) to express $\psi$ via $\varphi$. 

Asymptotic behavior of the SED shape function $\Phi(z)$ can be easily derived in the limit $z\ll 1$ [Eq. (\ref{eq:low_freq})] by expanding the argument of the exponential in the denominator of the equation (\ref{eq:low_freq}), and in the limit $z\gg 1$ [Eq. (\ref{eq:high_freq})] by noticing that the integral is dominated by $\xi\ll 1$ and setting $f(\xi)\to f(0)$. The corresponding behaviors are characterized by the constants 
\ba    
c_6(\lambda)&=&\frac{45}{\pi^4}\int\limits_0^\infty\left[f(\xi)\xi^5\right]^{1/4}d\xi,
\label{eq:c6}\\
%%%%%%%
c_7&=&\frac{180}{7\pi^4}\Gamma\left(\frac{16}{7}\right)\zeta\left(\frac{16}{7}\right)\approx 0.439782,
\label{eq:c7}
\ea    
where $\zeta(x)$ is a Riemann's $\zeta$-function \citep{Abram}. 

The high-frequency asymptotic changes in the case of a disk with no central torque ($\lambda=\lambda_0$), as then we cannot set $f(\xi)\to f(0)=0$. Instead, we use the fact that $f(\lambda_0,\xi)\to\lambda_0\xi$ as $\xi\to 0$ for $\lambda=\lambda_0$ and substitute this behavior in the integrand. As a result we arrive at the equation (\ref{eq:high_freq_no_torque}) with $\lambda$ given by equation (\ref{eq:lamLd}) and $c_8$ given by
\ba   
c_8=\frac{30}{\pi^4}\Gamma\left(\frac{8}{3}\right)\zeta\left(\frac{8}{3}\right)\approx 0.595066.
\label{eq:c8}
\ea   

%%%%%%%%%%%%%%%%%%%%%%%%%%%%%%%%%%%%%%%%%%%%%%%%%%%%%%%%%%%
%%%%%%%%%%%%%%%%%%%%%%%%%%%%%%%%%%%%%%%%%%%%%%%%%%%%%%%%%%%

\section{Details of the solution determination}  
\label{sect:sol_det}

%%%%%%%%%%%%%%%%%%%%%%%%%%%%%%%%%%%%%%%%%%%%%%%%%%%%%%%%%%%

Here we provide some details of the similarity solution determination covered in \S \ref{sect:sol_barrier}-\ref{sect:sol_torque}. In the case when $\lambda$ is fixed (e.g. through $\dot m$, see \S \ref{sect:sol_sup}), the knowledge of $M_d$ and $L_d$ at time $t=0$ allows one to find, using equations (\ref{eq:M1}) and (\ref{eq:L1}), that 
\ba    
F_0 &=& \left[D_{J,0}M_d^{2-p}\left(I_L^{-1}L_d(0)\right)^{p-1}\right]^{1/(1-d)},
\label{eq:F0}\\
%%%%%%%%%%%
l_0 &=& I_L^{-1}\frac{L_d(0)}{M_d(0)}.
\label{eq:l0}
\ea     
Note that the dependence on $\lambda$ enters only through $I_L(\lambda)$.

For the model, in which the physics of the central barrier is adequately characterized by the equation (\ref{eq:prescr1}), one finds
\ba   
F_0 &=& \left[D_{J,0}M_d(0)\left[\frac{K(f(\lambda,0))^\eta}{f^\prime_\xi(\lambda,0)}\right]^{1-p}\right]^{1/(\lambda(1-d))},
\label{eq:F0_phys}\\
%%%%%%%%%%%%%%%%
l_0 &=& F_0^{1-\eta}\frac{f^\prime_\xi(\lambda,0)}{K(f(\lambda,0))^\eta},
\label{eq:l0_phys}
\ea   
where $\lambda$ is given by equation (\ref{eq:lam_phys}) and $f(\lambda,0)$, $f^\prime_\xi(\lambda,0)$ are evaluated for this particular value of the similarity parameter.

\end{document}